%% file: Matching.tex
\documentclass[onecolumn,epjc3]{svjour3} 
 
\usepackage[T1]{fontenc}
\usepackage[utf8]{inputenc}
\usepackage{epsfig,amsmath,amssymb}
\usepackage{graphicx}
\usepackage{xspace}
\usepackage{listings}
\usepackage{ulem}
\usepackage{slashed}
\usepackage{cite}
\usepackage{nicefrac}
\usepackage{feynmp}
\usepackage[colorlinks,citecolor=blue,urlcolor=blue,linkcolor=blue]{hyperref}
\journalname{Eur. Phys. J. C}

\newcommand{\DR}{\ensuremath{\overline{\text{DR}}}\xspace}
\newcommand{\MS}{\ensuremath{\overline{\text{MS}}}\xspace}
\newcommand{\re}{\Re\text{e}}

\usepackage{textcomp}

\newcommand\SARAH{{\tt SARAH}\xspace}

\newcommand\SPheno{{\tt SPheno}\xspace}

\newcommand{\MSbar}{{\ensuremath{\overline{\mathrm{MS}}}}}

\newcommand{\newc}{\newcommand}

\newc{\Psibar}{\overline{\Psi}}
\newc{\FFbS}{\overline{FF}S}
\newc{\FFbV}{\overline{FF}V}
\newc{\FFbe}{\overline{FF}\epsilon}
\newc{\FSS}{F_{SS}}
\newc{\FSSS}{F_{SSS}}
\newc{\FFFS}{F_{FFS}}
\newc{\FFFbS}{F_{\overline{FF}S}}
\newc{\FSSV}{F_{SSV}}
\newc{\FVS}{F_{VS}}
\newc{\FVVS}{F_{VVS}}
\newc{\FFFV}{F_{FFV}}
\newc{\FFFbV}{F_{\overline{FF}V}}
\newc{\FVV}{F_{VV}}
\newc{\FVVV}{F_{VVV}}
\newc{\fggV}{f_{ggV}}
\newc{\Fgauge}{F_{\rm gauge}}
\newc{\fSS}{f_{SS}}
\newc{\fSSS}{f_{SSS}}
\newc{\fFFS}{f_{FFS}}
\newc{\fFFbS}{f_{\overline{FF}S}}
\newc{\fSSV}{f_{SSV}}
\newc{\fVVS}{f_{VVS}}
\newc{\fVS}{f_{VS}}
\newc{\fFFV}{f_{FFV}}
\newc{\fFFbV}{f_{\overline{FF}V}}
\newc{\fVV}{f_{VV}}
\newc{\fVVV}{f_{VVV}}
\newc{\fgauge}{f_{\rm gauge}}
\newc{\FFVbar}{{\overline{FF}V}}
\newc{\FFSbar}{\overline{FF}S}

\usepackage{listings}

\newcommand{\MZ}{m_Z}
\newc{\bsigmu}{\bar\sigma^\mu}

\def\twov[#1,#2]{\left( \begin{array}{c} #1 \\ #2 \end{array} \right)}
\def\twoa[#1,#2][#3,#4]{\left( \begin{array}{cc} #1 & #2 \\ #3 & #4 \end{array} \right)}

\begin{document}

\title{Improved predictions for intermediate and heavy Supersymmetry in the MSSM and beyond}
\author{
   Florian Staub \thanksref{a2,a3} \and
   Werner Porod \thanksref{a1}
   }

\institute{
Institute for Theoretical Physics (ITP), Karlsruhe Institute of Technology, Engesserstra{\ss}e 7, D-76128 Karlsruhe, Germany\label{a2}
\and
Institute for Nuclear Physics (IKP), Karlsruhe Institute of Technology, Hermann-von-Helmholtz-Platz 1, D-76344 Eggenstein-Leopoldshafen, Germany\label{a3}
\and
Institut f\"ur Theoretische Physik und Astronomie,Universit\"at W\"urzburg, Am Hubland, 97074 W\"urzburg, Germany\label{a1}
}

\date{KA-TP-03-2017}

\maketitle

\lstset{frame=shadowbox}
\lstset{prebreak=\raisebox{0ex}[0ex][0ex]
        {\ensuremath{\hookrightarrow}}}
\lstset{postbreak=\raisebox{0ex}[0ex][0ex]
        {\ensuremath{\hookleftarrow\space}}}
\lstset{breaklines=true, breakatwhitespace=true}
\lstset{numbers=left, numberstyle=\scriptsize}
 
\begin{abstract}
For a long time, the minimal supersymmetric standard model (MSSM) with light masses for the supersymmetric states was considered as the most natural extension of the Standard Model of particle physics. 
Consequently, a valid approximation was to match the MSSM to the precision measurement directly at the electroweak scale. This approach was also utilized by all dedicated spectrum
generators for the MSSM. However, the higher the supersymmetric (SUSY)
scale is, the bigger the uncertainties which are introduced by this matching. We point out important consequences of a
two-scale matching, where the running parameters within the SM are calculated at $M_Z$ and  evaluated up to the SUSY scale, where they are matched to the full model.  We show the impact 
on gauge coupling unification as well as the SUSY mass spectrum. Also the Higgs mass 
prediction for large supersymmetric masses has been improved by performing the calculation within an effective SM. The approach presented here is now also available in the spectrum generator \SPheno. Moreover, also \SARAH was extended 
accordingly and gives the possibility to study these effects now in many different supersymmetric models. 
\end{abstract}


\input{main}

\begin{appendix}
\input{appendix}
\end{appendix}

\bibliographystyle{ArXiv}
\bibliography{lit}

\end{document}

%% file: main.tex
\section{Introduction}
The discovery of the Higgs with a mass of about 125~GeV \cite{Aad:2012tfa,Chatrchyan:2012xdj} is, to date, the biggest 
success of the large hadron collider (LHC). In contrast, there has not been any evidence for new physics. This puts very 
strong constraints on the masses of new coloured particles as predicted, for instance, by supersymmetry (SUSY); working with 
very simplified assumptions, it is possible to exclude gluinos and first/second generation squarks nearly up to 2~TeV 
\cite{ATLAS-CONF-2016-052,ATLAS-CONF-2016-054,ATLAS-CONF-2016-078,Xie:2223502}. 
These experimental results raise not only the question if minimal supersymmetry is still a good solution to the fine-tuning 
or hierarchy problem of the standard model of particle physics (SM), but also gives new  challenges to study the MSSM 
precisely.

In the past many studies for the MSSM were done under the impression that the scale of supersymmetry, $M_{\rm SUSY}$,
should be close to the electroweak scale $M_Z$. Under this assumption it was
possible to calculate the gauge couplings in the \DR\ scheme directly from $\MZ$, $G_F$
and $\alpha_{em}$ as well as the \DR\ Yukawa couplings from the pole-mass of the top-quark
and the running \MSbar\ lepton and light quark masses given at $Q=\MZ$. More importantly
the Higgs mass(es) has been calculated at fixed order in the full supersymmetric model.
However, both calculations became less accurate the larger $M_{\rm SUSY}$ is because  potentially large logarithms of 
form $\log\frac{M_{\rm SUSY}}{M_Z}$ and $\log\frac{M_{\rm SUSY}}{m_h}$, respectively, appear.
Therefore, there are ongoing efforts 
to improve the calculation in the presence of supersymmetric scales which are well above the electroweak one. 
The first road is to keep the current set-up in principle but improve it by higher order corrections: for instance, 
{\tt SoftSUSY} provides the possibility to include higher order corrections to the threshold corrections at the weak 
scale and in the renormalisation group equation (RGE) running between the weak and SUSY scale, in order to get a better 
determination of the \DR parameters at the SUSY scale \cite{Allanach:2014nba}.  The first ansatz 
is to calculate the Higgs mass still in the full MSSM but extends the two-loop fixed order calculation by a resummation of potential large logarithm 
involving stops. That's for instance done by {\tt FeynHiggs} since a few years \cite{Hahn:2009zz,Heinemeyer:1998yj,Hahn:2013ria}. The second approach, which becomes more and more popular, is to work in an effective 
theory below $M_{\rm SUSY}$: {\tt SusyHD} \cite{Vega:2015fna} and recent versions of {\tt FlexibleSUSY} \cite{Athron:2016fuq} 
as well as {\tt FeynHiggs} \cite{Bahl:2016brp}
can consider below $M_{\rm SUSY}$ only the degrees of freedom of the SM, and match the SM to the MSSM just at the SUSY scale.  
Also the Higgs mass calculation is done in the effective SM by obtaining a value of the quartic Higgs coupling 
$\lambda_{\rm SM}$ from the matching between the MSSM and SM at $M_{\rm SUSY}$. The idea to work in an effective SM below 
$M_{\rm SUSY}$ was already well explored in literature before it became easily available via public tools, see e.g. 
Refs.~\cite{Espinosa:1991fc,ArkaniHamed:2004fb,Giudice:2004tc,Giudice:2011cg,Degrassi:2012ry,%
Draper:2013oza,Bagnaschi:2014rsa}.
Similarly, also a general Two-Higgs-Doublet-Model was already considered as low energy limit of the MSSM 
\cite{Haber:1993an,Carena:1995bx,Lee:2015uza}. 
Finally, since several years Split-SUSY variants of the MSSM become more and more popular in which the coloured SUSY 
particles are integrated out 
\cite{ArkaniHamed:2004fb, Giudice:2004tc, ArkaniHamed:2004yi,Kilian:2004uj,Bernal:2007uv, Giudice:2011cg}. 

We have now also extended the stand-alone spectrum generator \SPheno \cite{Porod:2003um,Porod:2011nf} as well as the
{\tt Mathematica} package \SARAH \cite{Staub:2008uz,Staub:2009bi,Staub:2010jh,Staub:2012pb,Staub:2013tta,Staub:2015kfa}, 
which gives the possibility to auto-generate a spectrum generator for a given model, to improve the predictions for 
moderate and heavy SUSY scales. Here, we made use of the second approach: the running \DR parameters at the SUSY scale are 
obtained via a two-scale matching procedure and the  Higgs mass calculation can optionally be done within an effective SM. We 
give in the following not only details of our exact approach but discuss also phenomenological consequences of the improved 
calculations. We focus not only on the Higgs mass prediction, which has been already discussed to some extent in the recent 
year, but show also potential important effects on the SUSY mass spectrum. Beside the MSSM we consider also
it is minimal extension, the NMSSM.

This paper is organised as follows: in sec.~\ref{sec:methods} we summarize our approach to obtain the \DR parameters
 at the SUSY scale as well as to calculate the mass of the SM-like Higgs. Many details for the matching are given in 
appendix~\ref{app:Matching}, where also the differences between stand-alone \SPheno and the \SARAH generated
version are discussed. 
In sec.~\ref{sec:results} we discuss the numerical impact of the improved calculation on the running parameters, 
but also on the SUSY and Higgs masses in the MSSM and beyond. We conclude in sec.~\ref{sec:conclusion}.

\section{Matching procedure and effective Higgs mass calculation}
\label{sec:methods}

\subsection{The two-scale matching in \SARAH}
So far, all dedicated MSSM spectrum generators such as {\tt SoftSUSY} 
\cite{Allanach:2001kg,Allanach:2009bv,Allanach:2013kza,Allanach:2014nba}, {\tt Suspect} \cite{Djouadi:2002ze} 
or \SPheno were adapting the 
procedure of Ref.~\cite{Pierce:1996zz} to obtain the running gauge and Yukawa couplings at the SUSY scale. All details 
of the calculations are summarised in Appendix \ref{app:OneScale}. The principle idea is that all measured SM parameters are 
already translated at $M_Z$ into \DR values taking into account the complete MSSM spectrum
which are then evaluated to the SUSY scale by using the RGEs of the MSSM. 
This procedure suffers from increasing uncertainties when the separation of the electroweak and SUSY scale becomes large. 
In order to reduce the theoretical uncertainty for 
large SUSY scales, {\tt SoftSUSY} is able since some time to include the two-loop SUSY thresholds in the calculation of the 
\DR parameters and to perform a three-loop RGE running between $M_Z$ and $M_{\rm SUSY}$. With these additional corrections, 
potential large effects in the prediction of the Higgsino mass parameter but also for the Higgs mass were found. The 
drawback of this ansatz is that it is computational very expensive and  slows down the evaluation 
of a given parameter point significantly. Moreover, only the effects of a more precise determination of the top Yukawa coupling on the 
Higgs mass are caught in this approach up to some extent, while still potential large logarithm in the fixed order Higgs mass calculation can be present. 

\begin{figure}[tb]
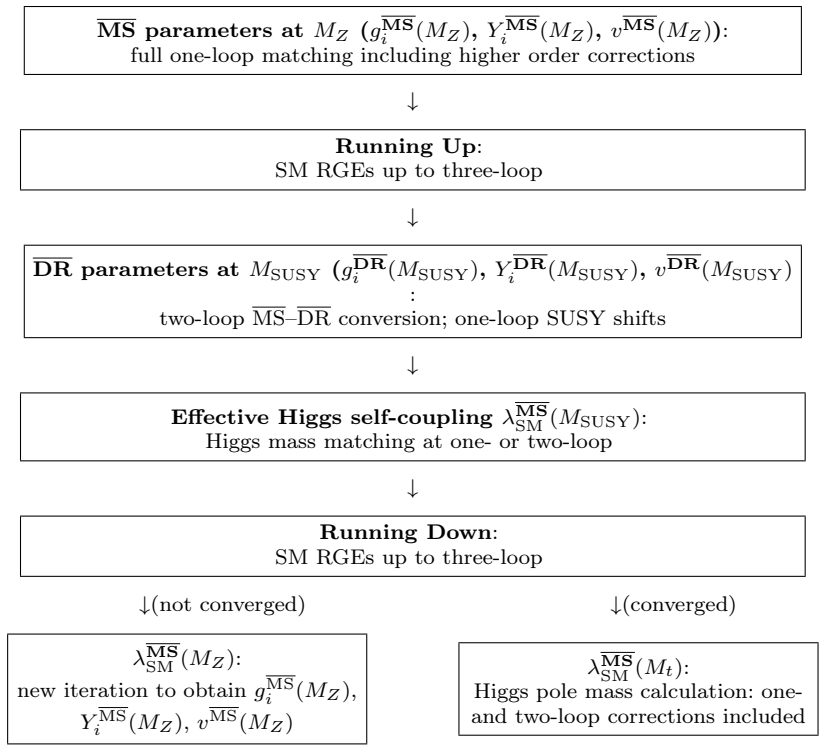

\begin{center}
\fbox{\parbox{10cm}{\centering {\bf \MS parameters  at $M_Z$ ($g_i^{\MS}(M_Z)$, $Y_i^{\MS}(M_Z)$, $v^{\MS}(M_Z)$)}: \\ full one-loop matching including higher order corrections}} \\[2mm]
\textdownarrow \\[2mm]
\fbox{\parbox{10cm}{\centering {\bf Running Up}: \\  SM RGEs up to three-loop } }\\[2mm]
\textdownarrow \\[2mm]
\fbox{\parbox{10cm}{\centering {\bf \DR parameters at $M_{\rm SUSY}$ ($g_i^{\DR}(M_{\rm SUSY})$, $Y_i^{\DR}(M_{\rm SUSY})$, $v^{\DR}(M_{\rm SUSY})$ }: \\ two-loop \MS--\DR conversion; one-loop SUSY shifts }}\\[2mm]
\textdownarrow \\[2mm]
\fbox{\parbox{10cm}{\centering {\bf Effective Higgs self-coupling $\lambda_{\rm SM}^{\MS}(M_{\rm SUSY})$}: \\  Higgs mass matching at one- or two-loop } }\\[2mm]
\textdownarrow \\[2mm]
\fbox{\parbox{10cm}{\centering {\bf Running Down}: \\  SM RGEs up to three-loop } }\\[2mm]
\hspace{2cm}\parbox{3cm}{\textdownarrow (not converged)} \hspace{3cm} \parbox{3cm}{\textdownarrow (converged)}\\[2mm]
\fbox{\parbox{4.5cm}{\centering {\bf $\lambda_{\rm SM}^{\MS}(M_Z)$}: \\ new iteration to obtain $g_i^{\MS}(M_Z)$, $Y_i^{\MS}(M_Z)$, $v^{\MS}(M_Z)$}}
\hspace{1cm}
\fbox{\parbox{4.5cm}{\centering {\bf $\lambda_{\rm SM}^{\MS}(M_t)$}: \\ Higgs pole mass calculation: one- and two-loop corrections included}}
\end{center}
\caption{Schematic procedure of the two scale matching.}
\label{fig:Matching}
\end{figure}

Therefore, we are using another ansatz in \SPheno and \SARAH\footnote{We use in the following \SARAH as synonym for 'a \SARAH generated spectrum generator based on \SPheno'} 
which is closer to the setup of {\tt NMSSMCalc} \cite{Baglio:2013iia} or specific versions 
of {\tt FlexibleSUSY} \cite{Athron:2014yba,Athron:2016fuq}: the matching at the electroweak scale includes only SM thresholds to obtain the \MS values of the 
gauge and Yukawa couplings and the electroweak vacuum expectation value (VEV). These parameters are then evolved up to the SUSY scale
using SM RGEs, and the translation from \MS to \DR scheme and the inclusion of SUSY thresholds is done at the SUSY scale. 
All details of the calculation are given in Appendix~\ref{app:Matching}. The precision to obtain the \DR parameters at the SUSY scale 
via this two-scale matching (2SM) is as follows in \SARAH/\SPheno
\begin{enumerate}
 \item The \MS parameters at the weak scale are calculated using:
  \begin{itemize}
   \item One-loop electroweak corrections to the fermion masses
   \item Two-loop QCD corrections to the top quark mass 
   \item One-loop corrections to $\delta_{VB}$ as well as one- and two-loop corrections to $\delta_\rho$
  \end{itemize}
 \item The SM RGEs are available up to three-loop
 \item The \MS--\DR conversion of the running fermion masses is done at two-loop $\alpha_s$ and at one-loop
    in case of the electroweak gauge couplings
 \item The \MS--\DR conversion of the gauge couplings is done at one-loop
 \item The SUSY thresholds are included at full one-loop
\end{enumerate}
The \DR parameters obtained in that way are then used to calculate the SUSY and Higgs masses at $M_{\rm SUSY}$. Since both, the matching at the $M_Z$ and 
$M_{\rm SUSY}$ depends on these masses, one needs to iterate the matching procedure. For this reason it is necessary to calculate 
the quartic self-coupling $\lambda_{\rm SM}(M_{\rm SUSY})$ within the SM which is a function of the SUSY masses and parameters. A handy and very general ansatz to 
obtain $\lambda_{\rm SM}(M_{\rm SUSY})$ was presented in Ref.~\cite{Athron:2016fuq}: one can match the Higgs pole masses in the full MSSM and the SM at the SUSY scale 
\begin{equation}
 m_h^{\rm SM,pole}(M_{\rm SUSY}) \equiv m_h^{\rm MSSM,pole}(M_{\rm SUSY})
\end{equation}
from what one can derive $\lambda_{\rm SM}$ as
\begin{equation}
(v^{\MS}(M_{\rm SUSY}))^2 \lambda_{\rm SM}(M_{\rm SUSY})  = (m_h^{\rm SM,tree}(M_{\rm SUSY}))^2 = (m_h^{\rm MSSM,pole}(M_{\rm SUSY}))^2 - \Pi_h(M_{\rm SUSY} )
\end{equation}
Here, $\Pi_h(M_{\rm SUSY} )$ are the radiative corrections to the Higgs mass within the SM which are calculated using \MS parameters at this scale, 
while the pole mass calculation in the MSSM involves \DR parameters. The equivalence of this ansatz to the matching of four point function as for instance performed 
in Refs.~\cite{Draper:2013oza,Bagnaschi:2014rsa} and used also by {\tt SUSYHD} has been explicitly shown in Ref.~\cite{Athron:2016fuq}.
SM RGEs are used are afterwards to run $\lambda_{\rm SM}$ to $M_Z$, and the \MS parameters are recalculated at this scale. This procedure is iterated until the mass spectrum at the SUSY scale has converged. 

\subsection{Differences between \SARAH and \SPheno in the new matching routines}
\label{sec:differences}

The above procedure corresponds to the details in \SARAH whereas the procedure implemented in the stand-alone
\SPheno differs in the following details:
\begin{itemize}
 \item at $Q=m_t$: the top Yukawa coupling is optionally replaced by the fit formula given by eq.~(57) of    Ref.~\cite{Buttazzo:2013uya}
 \item at $Q=m_t$: the strong coupling $g_3$ is optionally  replaced by the fit formula given  by eq.~(60) of Ref.~\cite{Buttazzo:2013uya}
 \item at $Q=M_{SUSY}$: the thresholds corrections to the gauge and Yukawa interactions are
    calculated in the electroweak basis assuming an unbroken $SU(2)_L \times U(1)_Y$.
    The full formulae are given in Appendix~\ref{app:Matching}.
\end{itemize}
The flags to use/not use the fit formulae of  Ref.~\cite{Buttazzo:2013uya} are given in Appendix~\ref{app:TwoScale}. If not indicated otherwise, 
these fit formulae are used in the following comparisons.

\subsection{The effective Higgs mass calculation}
So far, the mass calculation with \SPheno/\SARAH would have stopped after the conversion of the mass spectrum at $M_{\rm SUSY}$. However, this could lead to a large theoretical uncertainty 
in the Higgs mass prediction for large SUSY masses: the fixed order Higgs mass calculation as performed by \SPheno/\SARAH would become inaccurate because of the appearance of large 
logarithms $\sim \log(M_{\rm SUSY}/M_{ew})$. In order to cure this, one could do a resummation of these large logs. However, in our setup it 
is much easier to use the value $\lambda_{\rm SM}(M_{\rm SUSY})$, which is already known, and run it to the top mass scale. By this running all large logarithms get re-summed and one can then calculate $m_h$ at $m_t$ within the SM including radiative corrections. In \SARAH/\SPheno we include the full SM one-loop corrections as well as the two-loop corrections $O(\alpha_t(\alpha_s\alpha_t))$ to $m_h$. The schematic procedure for the matching and Higgs mass calculation is summarized in Fig.~\ref{fig:Matching}.

\section{Consequences of the two-scale matching \& effective Higgs mass calculation}
\label{sec:results}
\subsection{Running SM couplings}
\label{sec:Coup}
\begin{figure}[ht]
\includegraphics[width=0.49\linewidth]{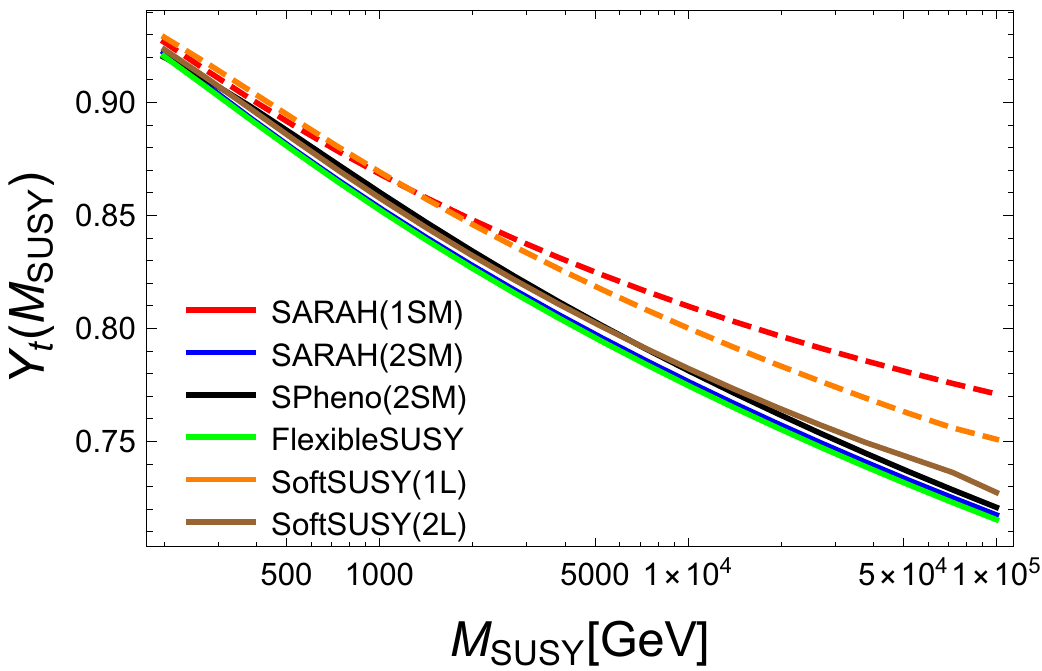} \hfill
\includegraphics[width=0.49\linewidth]{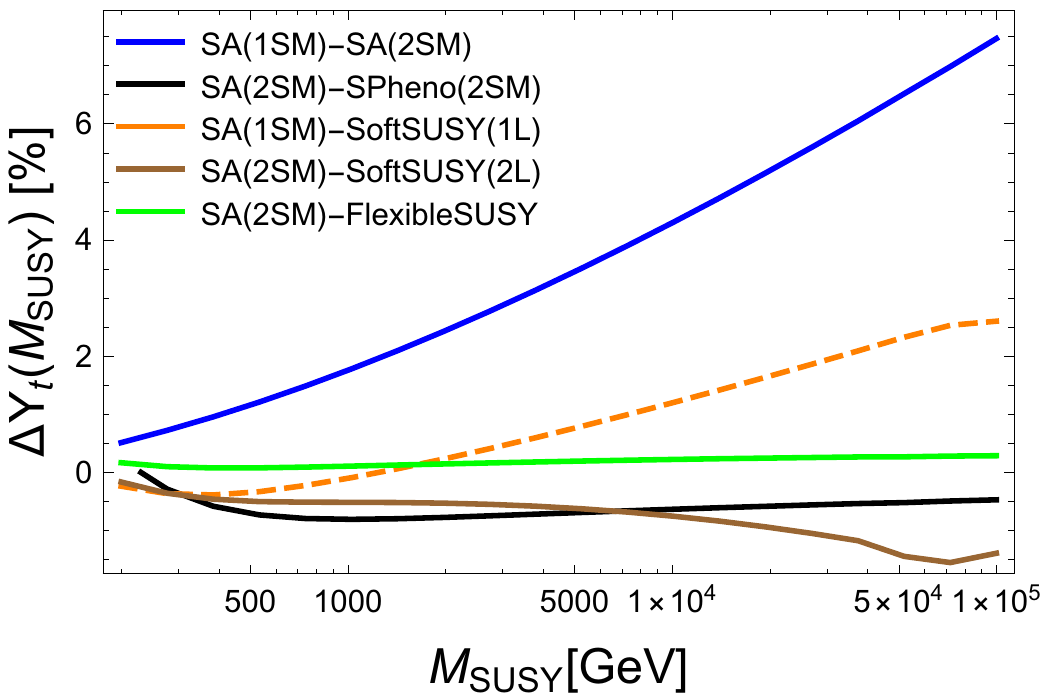} \\
\includegraphics[width=0.49\linewidth]{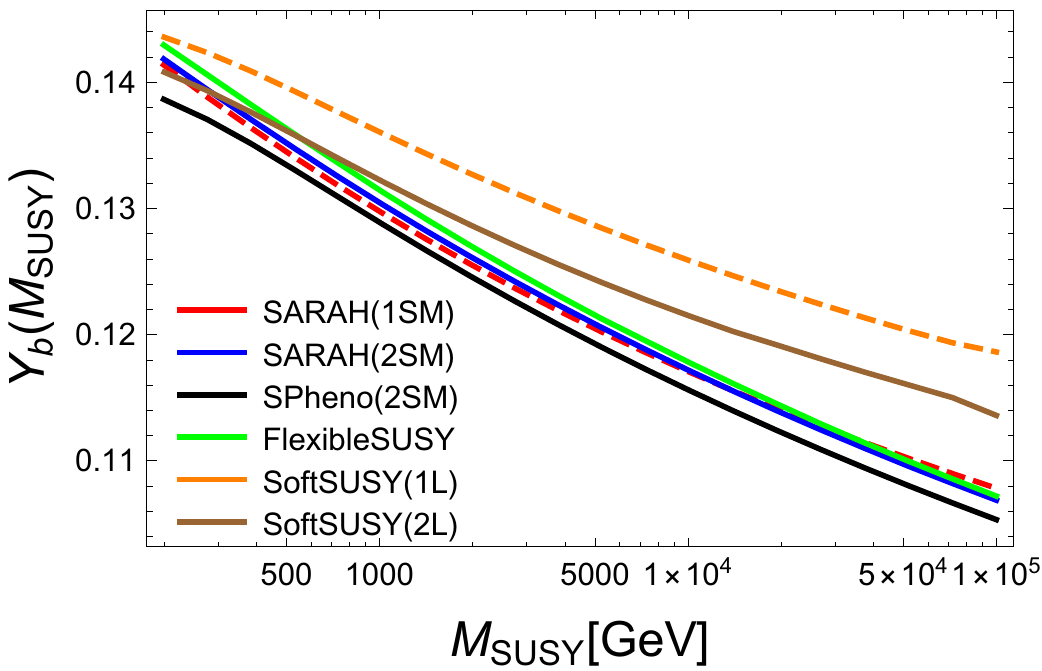} \hfill
\includegraphics[width=0.49\linewidth]{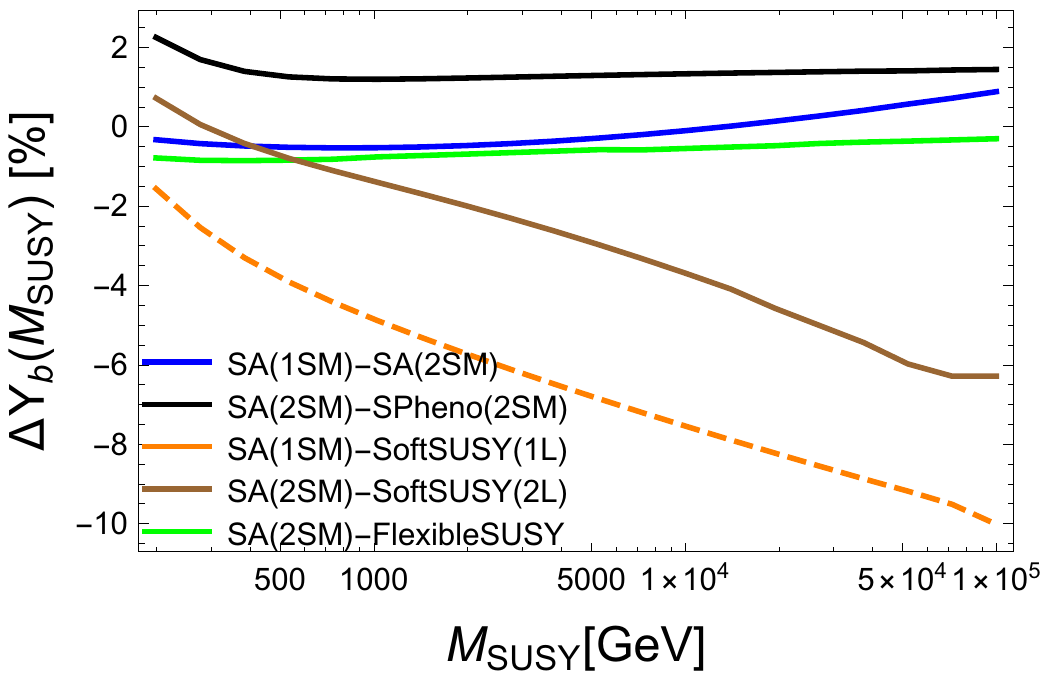} 
\caption{The \DR values of the running top and bottom
Yukawa couplings at the SUSY scale. The dashed red line shows the result 
using the one-scale matching as done by earlier \SARAH/\SPheno version, while the blue line is the new results from \SARAH and 
black the one from \SPheno. In addition, we show the results for {\tt SoftSUSY} using one-loop (dashed orange) and two-loop SUSY thresholds (full brown),
as well as for {\tt FlexibleSUSY} (green). 
On the right we give the difference $\Delta = \frac{Y^{\rm A} - Y^{\rm B}}{Y^{\rm A}}$ between 
the results of two calculations as indicated.}
\label{fig:Yukawas}
\end{figure}
\begin{figure}[ht]
\includegraphics[width=0.49\linewidth]{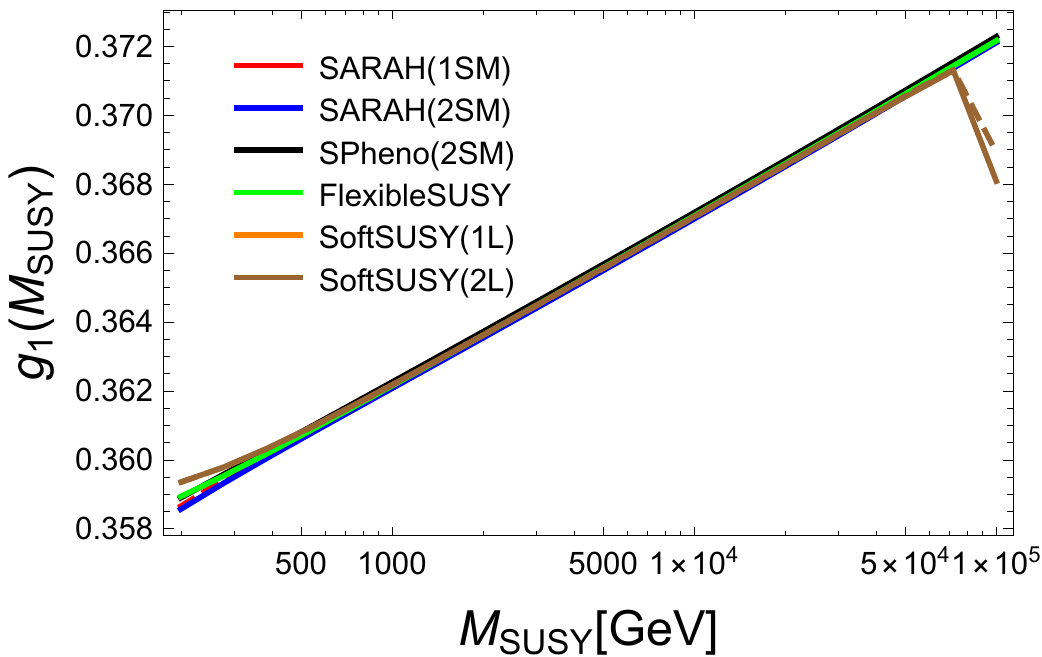} \hfill
\includegraphics[width=0.49\linewidth]{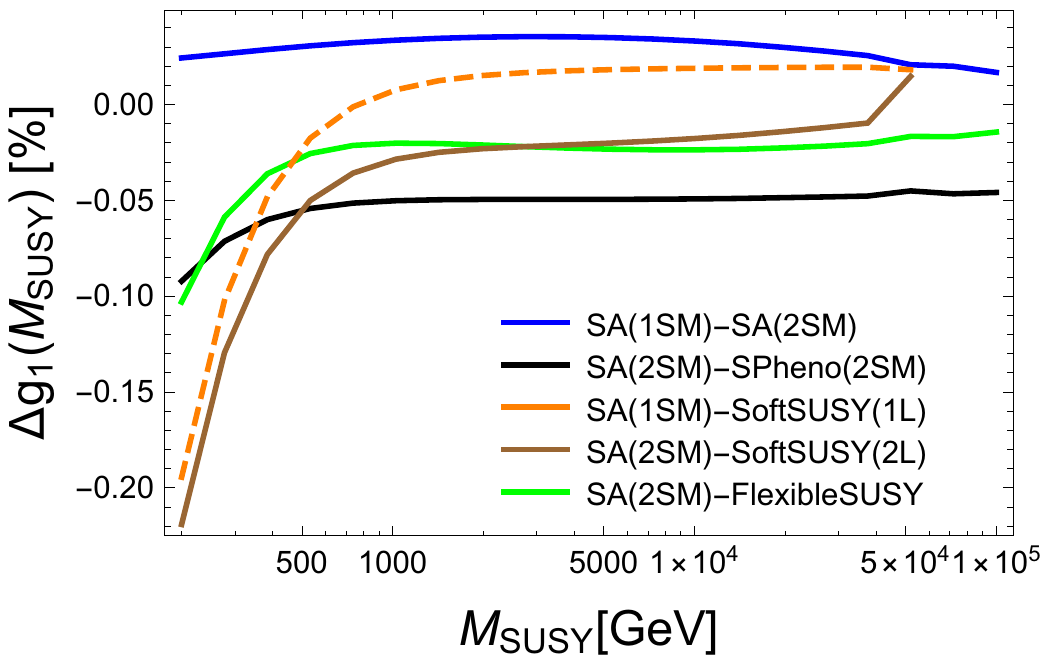} 
\includegraphics[width=0.49\linewidth]{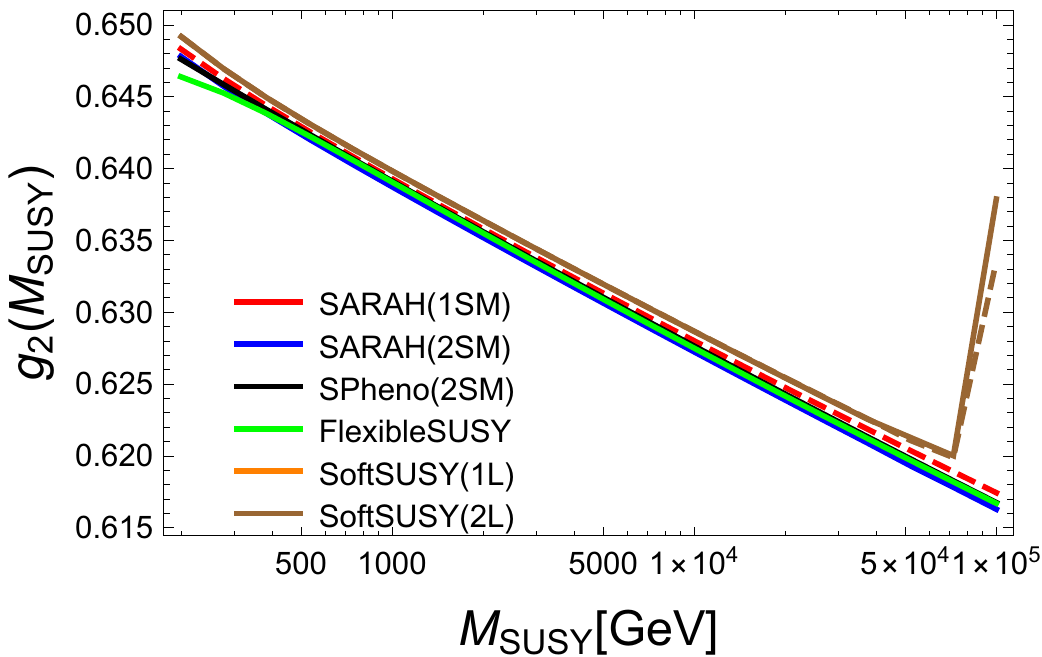} \hfill
\includegraphics[width=0.49\linewidth]{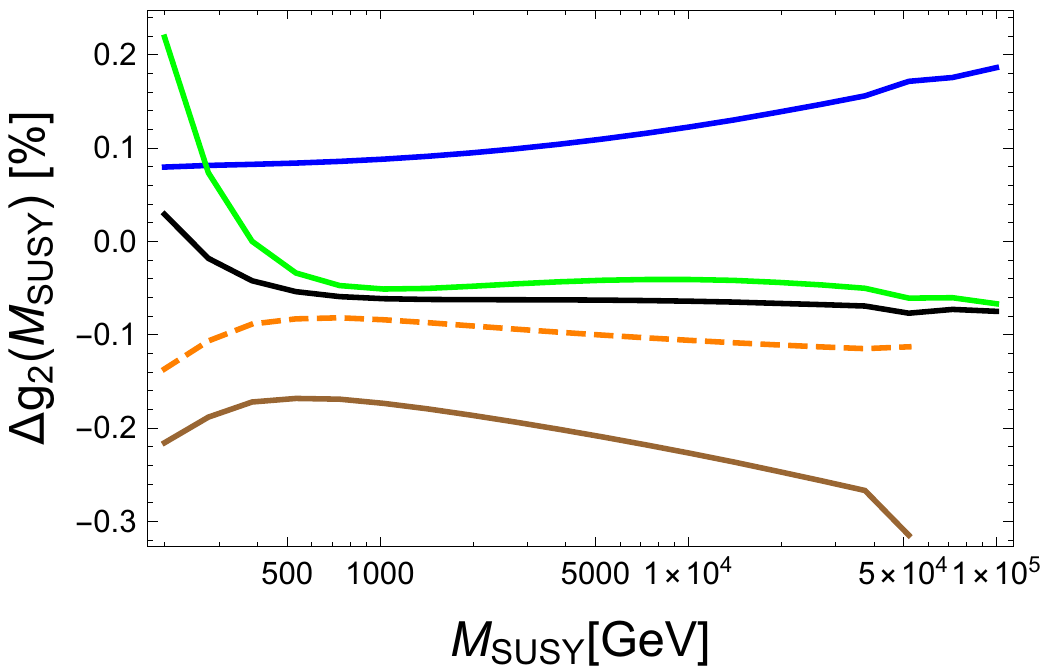} 
\includegraphics[width=0.49\linewidth]{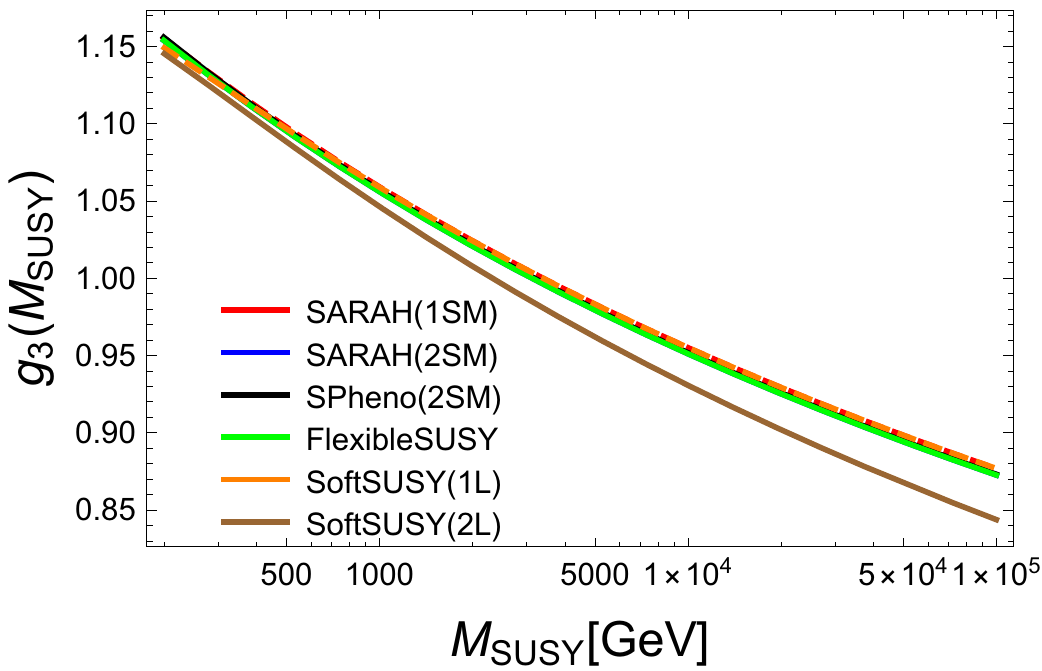} \hfill
\includegraphics[width=0.49\linewidth]{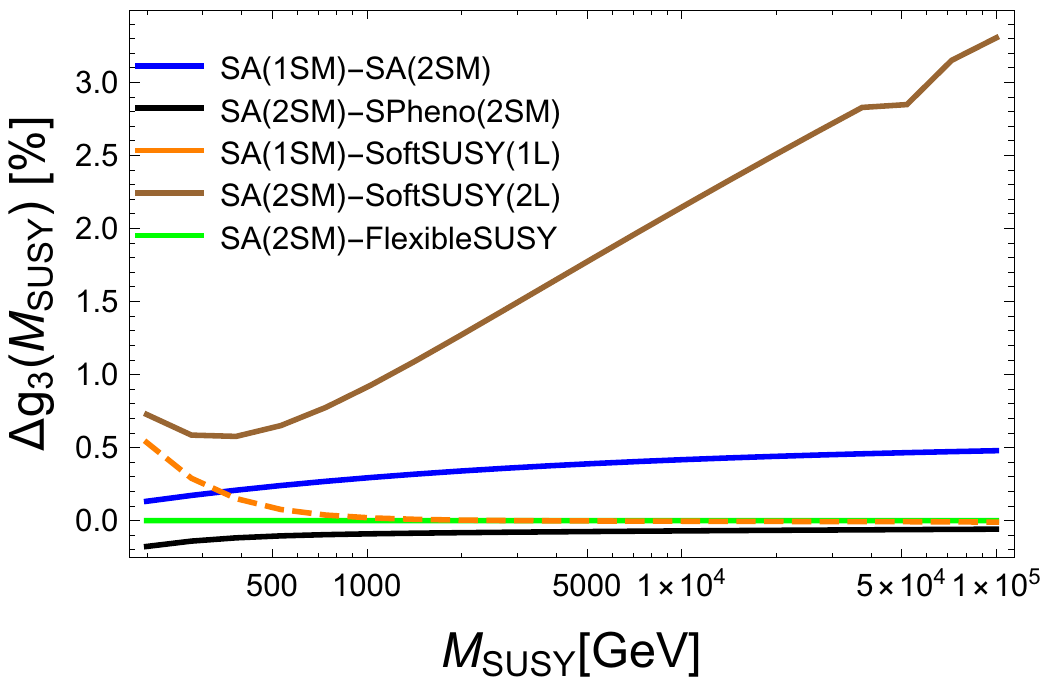} \\
\caption{The same as Fig.~\ref{fig:Yukawas} for the gauge couplings $g_1$, $g_2$ and $g_3$.}
\label{fig:gauge}
\end{figure}

All the efforts to disentangle the weak and the SUSY scale in the matching are done to get more accurate values of the running \DR parameters at the SUSY scale. Therefore, we want 
to start the discussion of the impact of the new matching procedure with presenting the changes in the 
\DR parameters at the SUSY scale. The results for the top and bottom Yukawa couplings 
are shown in Fig.~\ref{fig:Yukawas} and those for the three gauge couplings $g_1$, $g_2$ and $g_3$ are depicted in Fig.~\ref{fig:gauge}. Since the exact matching procedure using two scales is slightly different between \SPheno and \SARAH as explained in sec.~\ref{sec:differences} we show the new results for both codes. Since we have turned off here the fit formula of Ref.~\cite{Buttazzo:2013uya} in the \SPheno calculation, the remaining differences appearing here are due to the threshold corrections of the gauge and Yukawa couplings at $M_{\rm SUSY}$.
One sees that in particular the top Yukawa coupling changes significantly compared to the older calculation with only one matching scale (1SM). For $M_{\rm SUSY} =100$~TeV, the calculated \DR value with \SARAH using the two-scale matching is nearly 10\% below the one for the one-scale matching.  These large changes are in agreement with the results of 
Ref.~\cite{Athron:2016fuq} where the impact of a 2SM on the top Yukawa coupling has also been analysed analytically.
We show for comparison also the calculated couplings in {\tt SoftSUSY} with and without two-loop SUSY thresholds and three loop RGEs. It is obvious that there was a non-negligible difference between the old results 
and the one-loop results of {\tt SoftSUSY} although both calculations were of the same order in perturbation theory. The reason are the matching conditions which 
can schematically be written as 
\begin{equation}
\label{mfermion}
     m_t^{\DR} = m_t^{pole} + \hat{m}_t \,\Sigma(\hat{m}_t^2) \;, 
\end{equation}
where all loop corrections are summarised in $\Sigma$. \SPheno uses $\hat{m}_t = m_t^{\DR}$ while {\tt SoftSUSY} and other codes set 
$\hat{m}_t = m_t^{pole}$. The result obtained with the new two-scale matching agrees now rather well with the {\tt SoftSUSY} results once 
the two-loop SUSY corrections in the matching are included up to several TeV. However, for even higher SUSY scales one finds that even the SUSY calculation with two-loop thresholds gives sizeable differences to the RGE re-summed calculation.  On the other side, we find an excellent agreement with {\tt FlexibleSUSY} which performs also a two-scale matching, but uses a different matching procedure at the SUSY scale \footnote{We have adapted the approach of Ref.~\cite{Pierce:1996zz} to a two scale approach: we calculate the \DR gauge and Yukawa couplings from the running \MS values of $\alpha_{ew}$, $\sin\Theta_W$, $g_3$ as well as from the running fermion masses and CKM matrix at the SUSY scale. The calculation is similar to the corresponding matching of the measured values of these parameters to the \DR parameters at $M_Z$ done before. All details are given in appendix~\ref{app:TwoScale}. In contrast, {\tt FlexibleSUSY} demands the 
equality of pole masses in the SM and MSSM at the SUSY scale to get the matching conditions for the SM gauge and Yukawa couplings.}.  
A similar, but less pronounced effect can be seen for the bottom Yukawa coupling. Here, the changes between the one and two-scale matching account for a shift of about 6\% for a SUSY scale of 
100~TeV.

For the gauge couplings, the difference between the one and two-scale matching are in general much smaller than for the Yukawa couplings. The changes are usually well below 1~\% even for a SUSY scale of 100~TeV. The only exception is {\tt SoftSUSY} when turning on the two-loop thresholds to the strong coupling. In that case a significant decrease in $g_3$ with increasing $M_{SUSY}$ is seen. This effect is not confirmed by the RGE re-summed calculations. 

\subsection{Gauge coupling unification}
\begin{figure}[hbt]
\includegraphics[width=0.5\linewidth]{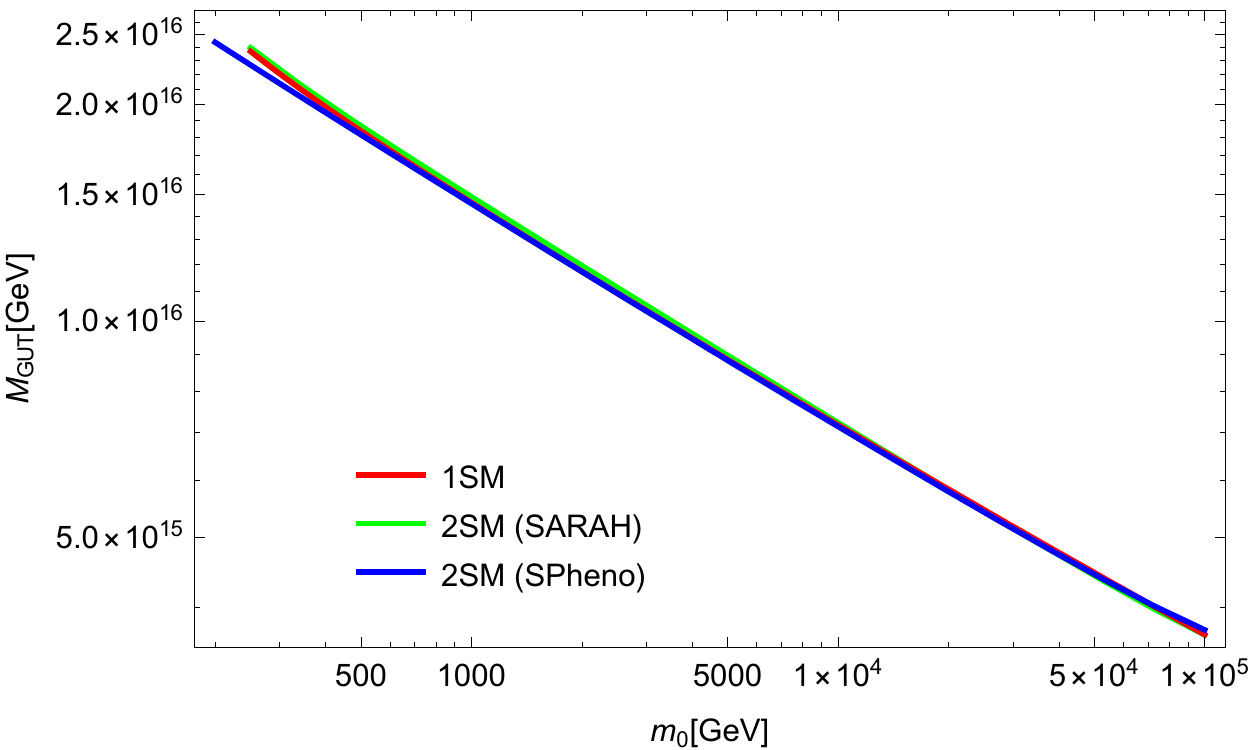} \hfill 
\includegraphics[width=0.5\linewidth]{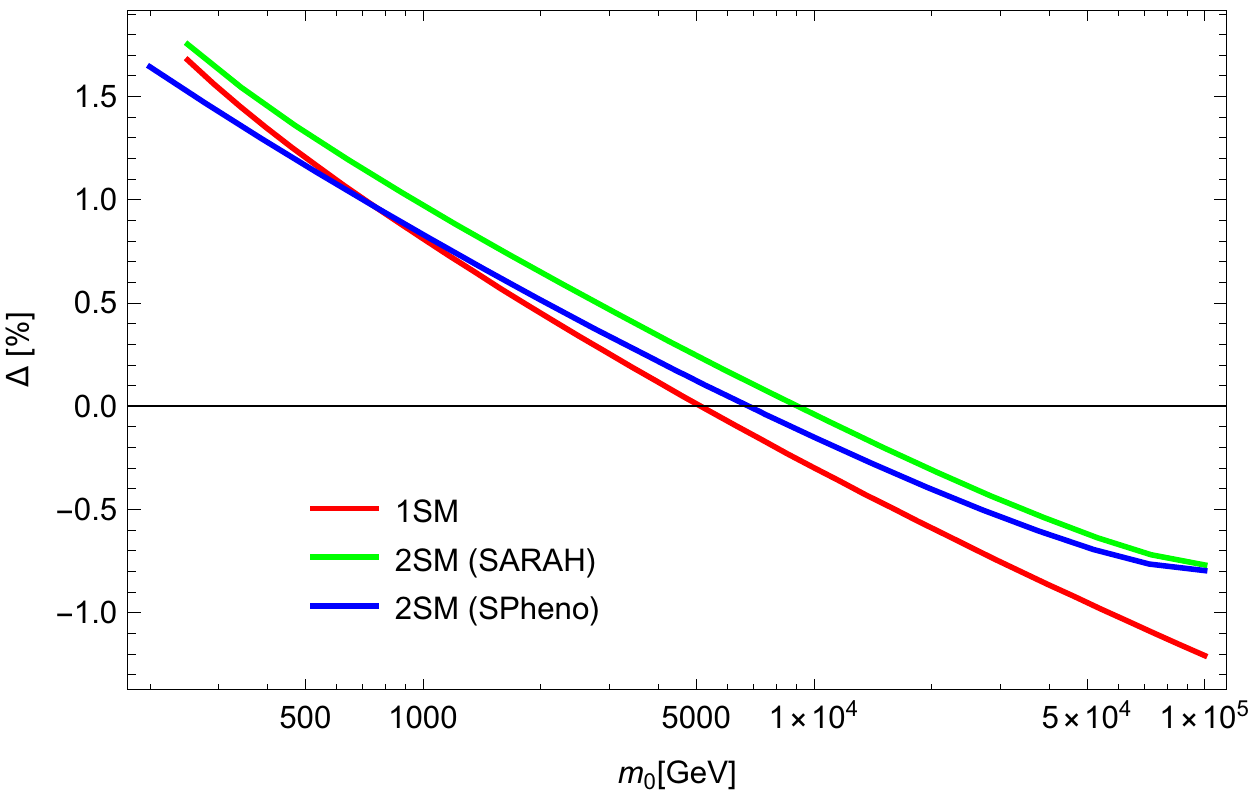}
\caption{On the left: the predicted value for $M_{\rm GUT}$ as function of $m_0=M_{1/2}$ in the CMSSM. The red line corresponds to the old one-scale matching, while the 
blue and green line are the results for the two-scale matching with \SARAH and \SPheno. On the right: the difference between $g_1(M_{\rm GUT})$ and $g_3(M_{\rm GUT})$ (in percent) 
as function of $m_0$. The colour code is the same as on the left. We included here in \SPheno the two-loop thresholds corrections to $g_3$. }
\label{fig:GUT}
\end{figure}

The shifts in the gauge couplings are rather small even for very large SUSY masses in the multi TeV range. Thus, they play phenomenologically only a sub-dominant role 
compared to the larger effects in the top Yukawa coupling. However, if one embeds the MSSM into a UV complete framework like supergravity, the running gauge couplings 
$g_1^{\DR}$ and $g_2^{\DR}$ are usually used as starting point to find the scale of grand unification, $M_{\rm GUT}$ by imposing the condition 
\begin{equation}
\label{eq:gGUT}
 g_1(M_{\rm GUT}) =  g_2(M_{\rm GUT})
\end{equation}
Also the goodness of complete unification, i.e. the remaining difference between $g_3(M_{\rm GUT})$ compared to the other two  couplings is very sensitive to the values of 
$g_1^{\DR}$ and $g_2^{\DR}$ at $M_{\rm SUSY}$. Therefore, we are checking the impact of the two-scale matching on $M_{\rm GUT}$ and $\Delta g = \frac{g_1(M_{\rm GUT})-g_3(M_{\rm GUT})}{g_1(M_{\rm GUT})}$
in a constrained version of the MSSM (CMSSM). The CMSSM
has five input the parameters: the universal scalar mass $m_0$, the universal gaugino mass $M_{1/2}$, the universal trilinear 
soft-breaking parameter $A_0$, the ratio of the ew VEVs $\tan\beta=v_u/v_d$ and the phase of $\mu$. All three dimensionful parameters, 
$m_0$, $M_{1/2}$ and $A_0$, are set $M_{\rm GUT}$. 
Here, we fixed 
\begin{equation}
m_0 = M_{1/2} \,,\hspace{1cm} \tan\beta = 10 \,,\hspace{1cm} A_0 = 0 \,,\mu >0
\end{equation}
and varied $m_0$ from 200~GeV up to 100~TeV. 
The results are shown in Fig.~\ref{fig:GUT}. The predicted value for the GUT scale as function of $M_{\rm SUSY}$ changes only slightly when using the new two-scale matching compared to the one-scale matching. In a complete GUT-model, the difference $\Delta g$ has to be explained by
threshold corrections to heavy GUT-scale particles \cite{Weinberg:1980wa,Hall:1980kf}
as we are using two-loop RGE running . Therefore, the right plot of this figure indicates the possible
size of such corrections due to the GUT-scale spectrum. The prediction for $\Delta g$ is different comparing the one- and two-scale matching, but also comparing the new results of \SARAH and \SPheno. The dominant origin of this difference is the inclusion of the  two-loop correction to $g_3$ in \SPheno, i.e. the difference between the two lines can be taken as an estimate for the theoretical uncertainty in $\Delta g$ coming from higher order effects: only two-loop SM corrections 
in the matching of $g_3$ are included in \SPheno, but not the two-loop SUSY thresholds. Also, for consistency three-loop RGEs 
of $g_3$ up to $M_{\rm GUT}$ would be necessary. However, for small $m_0$ also the terms 
$O(v^2/M^2_{\rm SUSY})$, which are neglected in \SPheno by computing the thresholds in 
the $SU(2)_L\times U(1)_Y$ limit become important and introduce a difference in the prediction of the GUT scale, which enters logarithmically in the unification condition. 

\subsection{SUSY masses}
\begin{figure}[ht]
\includegraphics[width=0.49\linewidth]{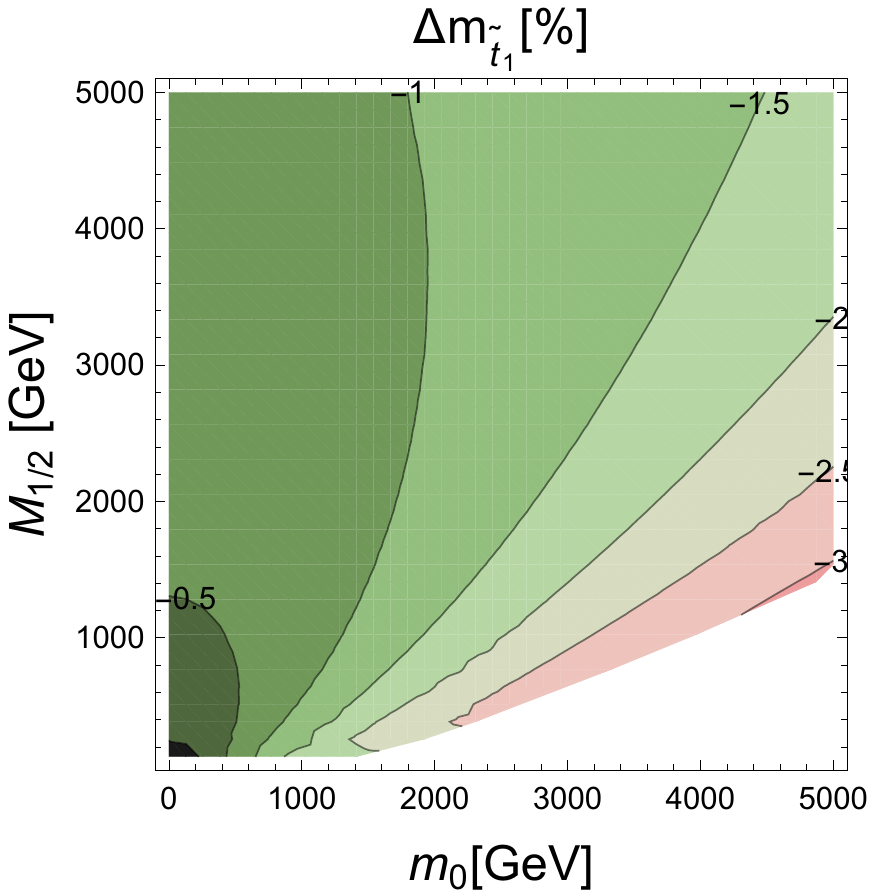} \hfill
\includegraphics[width=0.49\linewidth]{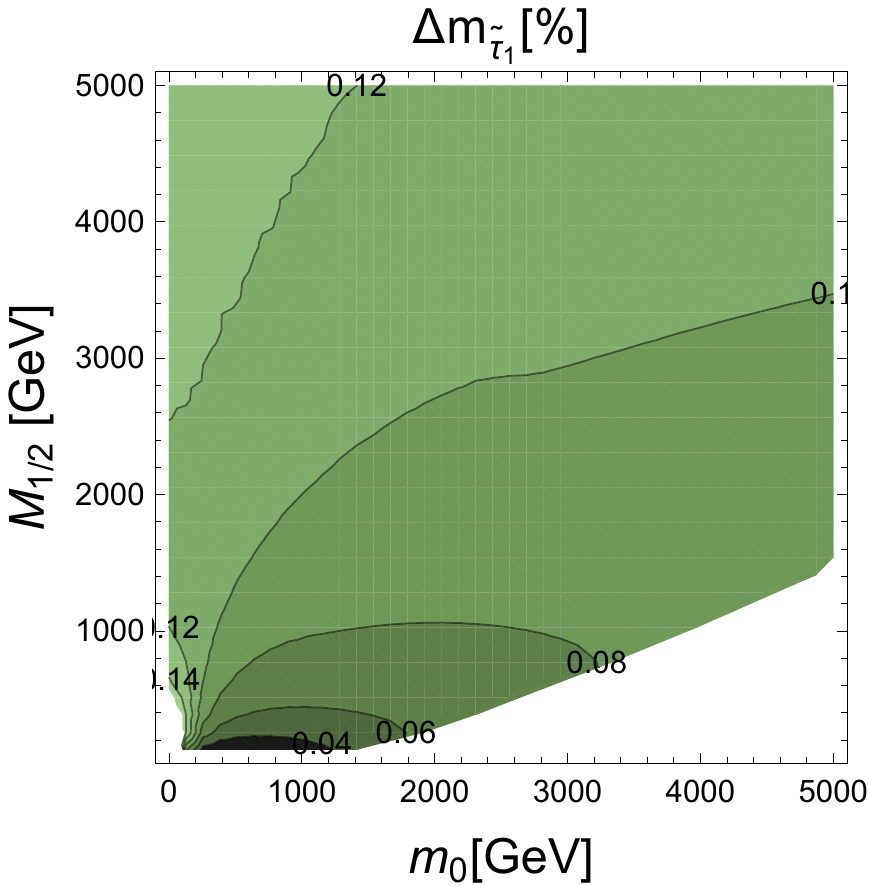} \\[5mm]
\includegraphics[width=0.49\linewidth]{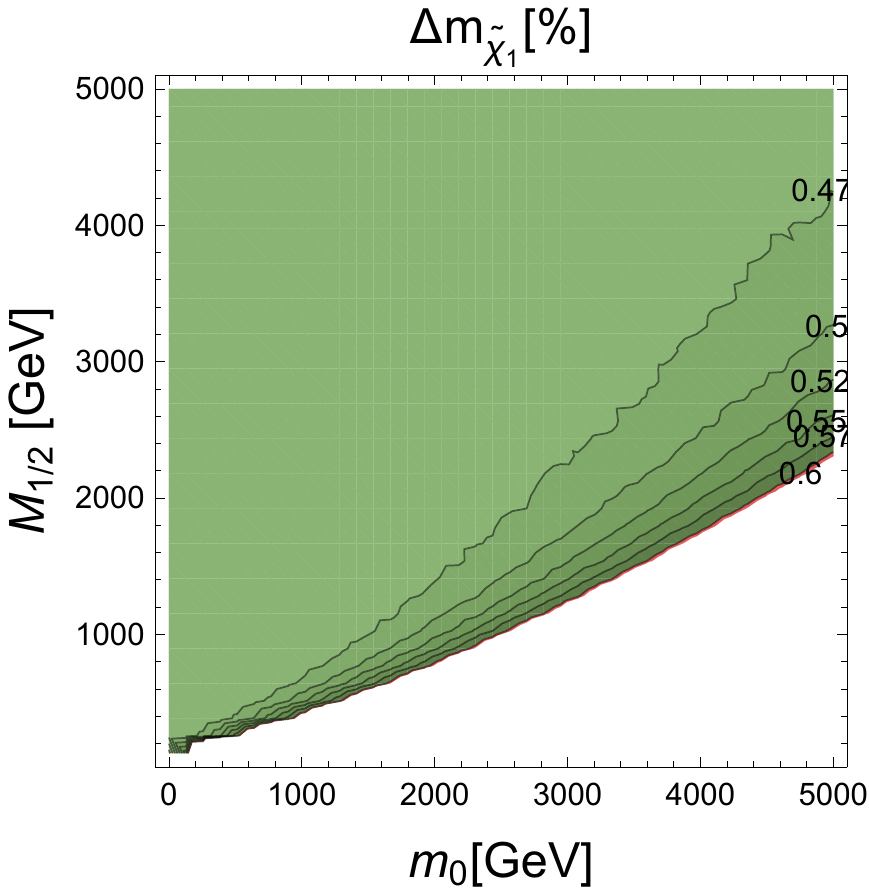} \hfill
\includegraphics[width=0.49\linewidth]{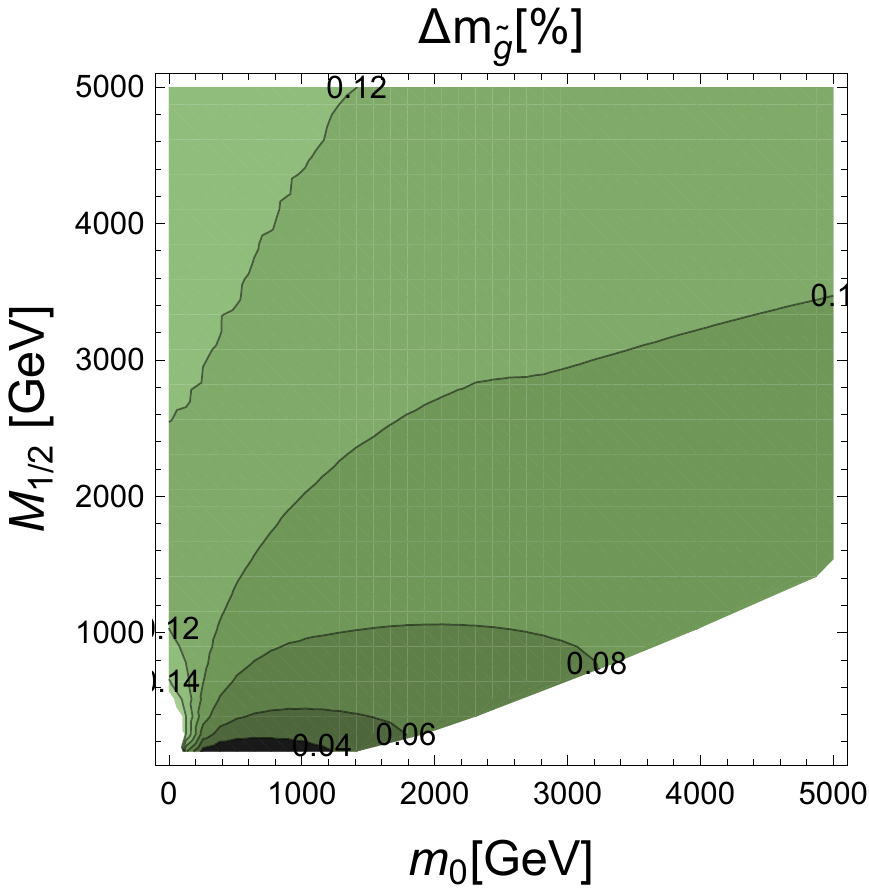} 
\caption{The mass difference $\Delta = \frac{m^{\rm old} - m^{\rm new}}{m^{\rm old}}$ in percent between the old and new mass calculation using  \SARAH. 
The red boundary in the $\tilde{\chi}^0_1$-plot shows
the area with a Higgsino LSP which is discussed in the text in more detail.}
\label{fig:masses}
\end{figure}

The changes in the \DR parameters at the SUSY scale influence also the mass spectrum. This has very important consequences 
in particular on the Higgs mass which are discussed in the dedicated section sec.~\ref{sec:mh}. 
For now, we concentrate on the 
SUSY masses. In that case, the masses do hardly change if all SUSY specific parameters are defined at the SUSY scale because only tiny 
changes in the $F$- and $D$-term contributions as well as in the radiative corrections will appear. Those are found to be hardly in the percent range even 
for large SUSY scales. Larger effects 
are present, if on considers unified scenarios in which the SUSY parameters are set via boundary conditions at a scale well above the 
SUSY scale. The additional RGE running between the high scale, which is often associated with the GUT scale via eq.~\ref{eq:gGUT}, 
will then introduce a larger dependence on \DR values of SM gauge and Yukawa couplings at $M_{\rm SUSY}$. As example, we consider again the CMSSM. 
For simplicity, we fix in the following, if not stated otherwise, $A_0 = 0$, $\mu > 0$, $\tan\beta = 10$ and perform a scan over $m_0$ and $M_{1/2}$. The changes 
in the masses of the lightest stop, lightest stau, lightest neutralino and the gluino in the $(m_0,M_{1/2})$-plane are shown in Fig.~\ref{fig:masses}. The 
largest effect in general can be seen for the light stop mass which changes by 2--3\% when pushing $m_0$ in the multi-TeV range. For the other masses, 
the changes in the \DR parameters account only for moderate changes of 1\% and below. 
The only exception are fine-tuned region with a Higgsino LSP 
which we discuss below in more detail. Here,
we also display the
changes in the bino LSP mass because there small shifts can have sizeable effects in the calculation of the relic
density, e.g.\ in case of Higgs resonances or in case of co-annihiliation.

\begin{figure}[hbt]
\includegraphics[width=0.49\linewidth]{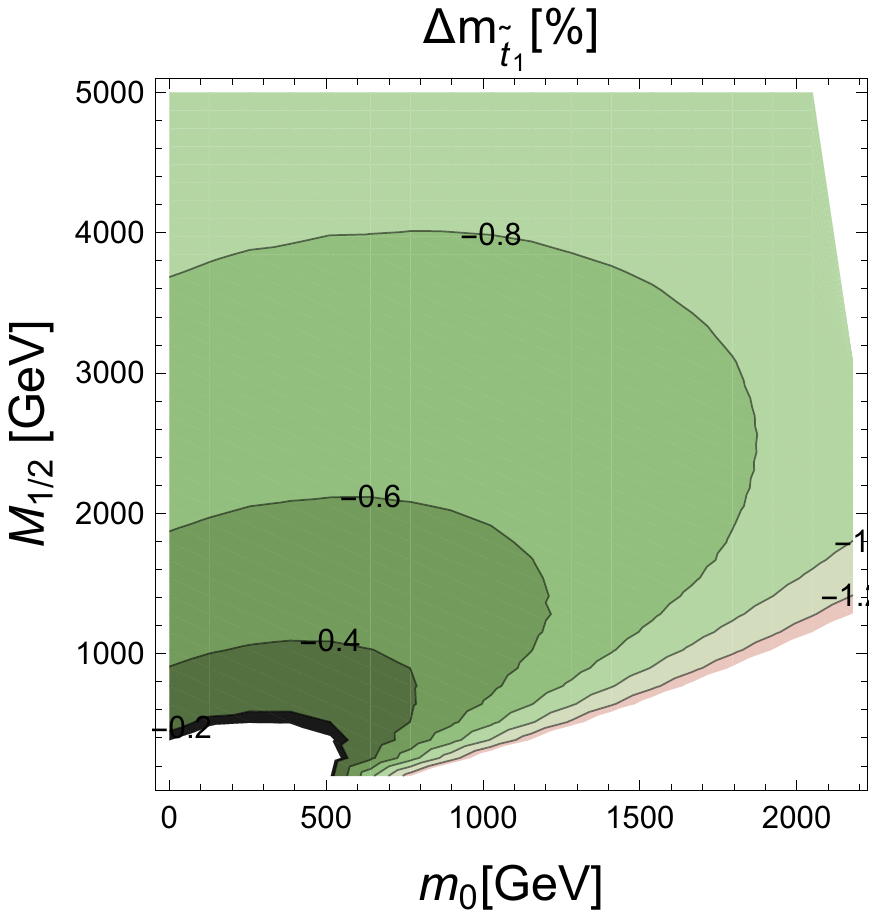} \hfill
\includegraphics[width=0.49\linewidth]{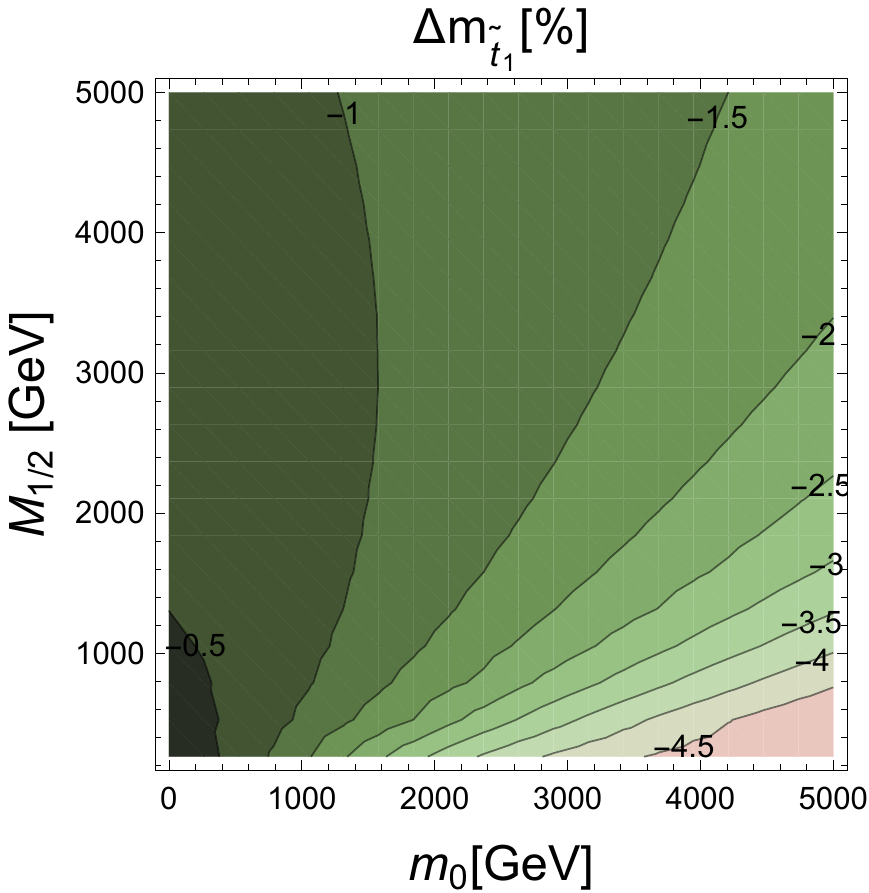} 
\caption{The same as Fig.~\ref{fig:masses} for lightest stop but using $A_0 = -1.5 m_0$ (left) and $A_0 = +1.5 m_0$ (right) .}
\label{fig:mSt1_A0}
\end{figure}

The impact of the \DR parameters at $M_{\rm SUSY}$ on the prediction of the light stop mass depends also on the chosen value for $A_0$. For non-vanishing $A_0$, 
the changes can become larger as shown in Fig.~\ref{fig:mSt1_A0}. Setting $A_0 = + 1.5 m_0$ we find that the stop mass changes by more than 5\% for $m_0 > 4$~TeV. 
These changes are still very moderate and have hardly any phenomenological impact at the LHC. 
However, as mentioned above they can become important for instance in stau or stop co-annihilation
 to explain the dark matter abundance in the universe \cite{Bagnaschi:2015eha}. 

\begin{figure}[hbt]
\includegraphics[width=0.49\linewidth]{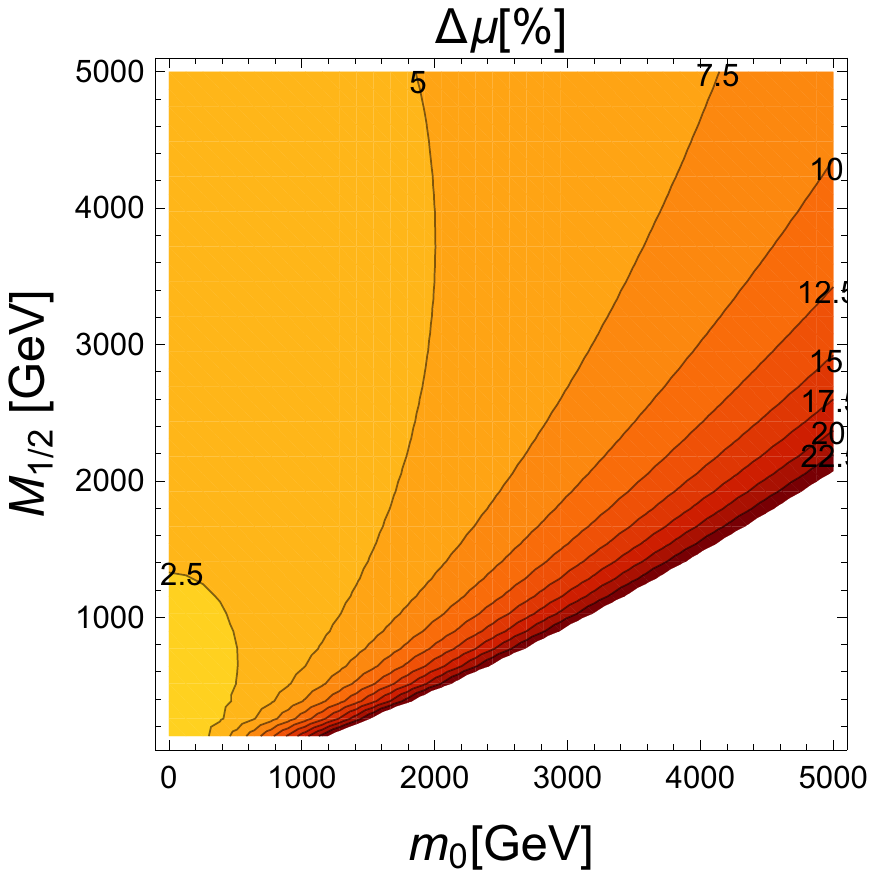} \hfill
\includegraphics[width=0.49\linewidth]{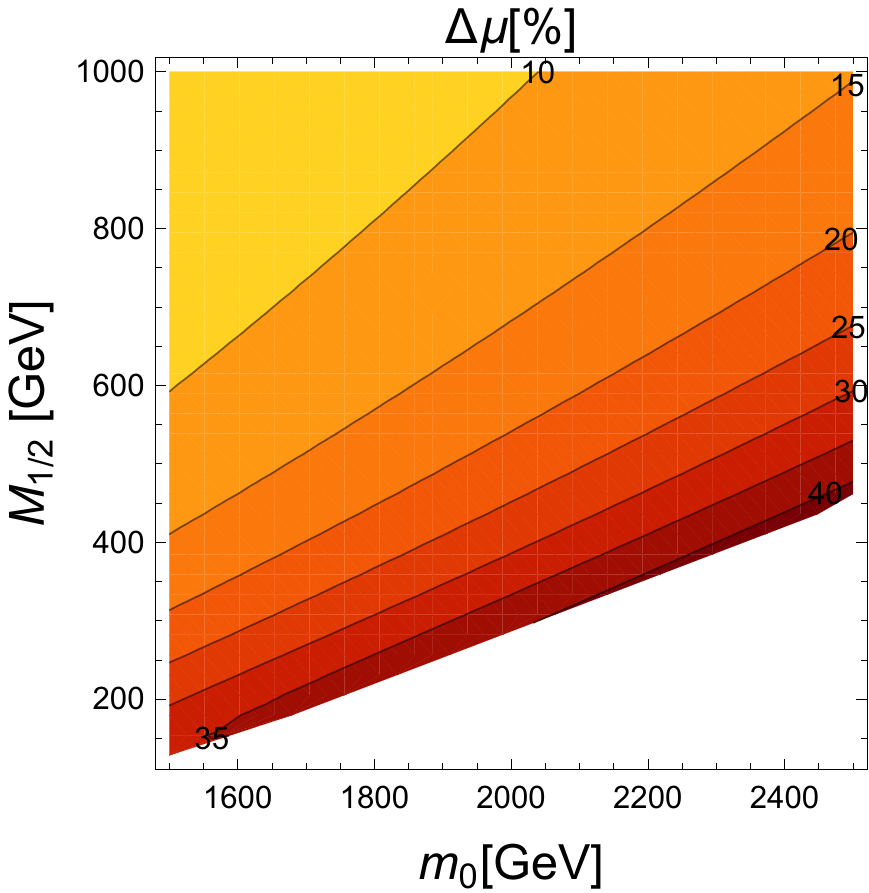} 
\caption{The same as Fig.~\ref{fig:masses} for the value of $\mu$ at the SUSY scale. The right plot is a zoom into the interesting region of the left one.}
\label{fig:mu}
\end{figure}

A much more pronounced effect can be observed for the $\mu$ parameter in the so called 'Focus-Point'-region \cite{Chan:1997bi,Feng:1999mn,Feng:1999zg,Feng:2000gh} from the minimisation conditions of the potential. This result at tree-level in
\begin{equation}
|\mu|^2 = \frac{(m_{H_d}^2 - m_{H_u}^2 \tan\beta^2)}{\tan^2\beta-1} - \frac{1}{2} M_Z^2 \simeq  - m_{H_u}^2  - \frac{1}{2} M_Z^2 
\end{equation}
where we have assumed in the last step $\tan\beta \gg 1$. The special feature of the focus point region is that 
cancellations in the RGE contributions 
to $m_{H_u}^2$ result in moderately small $\mu$ which is much smaller than the other SUSY mass parameters. How well these cancellation work 
depends strongly on the value of the top Yukawa coupling. Hence, we find that in the focus point region, which is usually needs moderate $M_{1/2}$ and 
large $m_0$, the value of $\mu$ changes by more than 25\% as shown in Fig.~\ref{fig:mu}. 
Thus, also the Higgsino masses vary significantly between the one and two-scale matching calculation.

If one assumes that a large $\mu$-parameter is the main source of fine-tuning in the MSSM, these changes in $\mu$ have also an impact on naturalness considerations. Using the 
approximate formula $\Delta \simeq \frac{\mu^2}{M_Z^2}$ as measure for the fine-tuning\footnote{These formula differs by a factor of two compared to the usually taken expression 
$\Delta \simeq 2 \frac{\mu^2}{M_Z^2}$ because of the incorporation of loop effects which have been overlooked for a long time \cite{Ross:2017kjc}.}, on sees that the fine-tuning prediction could reduce a factor of 2 and more in the focus point region when going from the one-scale matching to the two-scale matching. 

\subsection{Higgs mass in the MSSM}
\label{sec:mh}
The impact of heavy SUSY masses on the Higgs mass is nowadays a widely discussed topic. 
While fixed order calculations suffer from increasing uncertainties, there are two 
methods to improve the accuracy: (i) resumming the stop contributions as done 
by {\tt FeynHiggs}; (ii) working with a EFT ansatz as first done by {\tt SusyHD}
and later incorporated in {\tt FlexibleSUSY} as well. The pole mass matching 
described in sec.~\ref{sec:methods}, which was used so far only in {\tt FlexibleSUSY}  and now 
also by \SPheno/\SARAH, has the additional
advantage that it includes terms $O(v^2/M_{\rm SUSY}^2)$. This is in contrast to previous calculations 
to obtain $\lambda_{\rm SM}$ from the effective potential which are used by {\tt SusyHD} for instance.
Thus, these EFT calculation have a larger uncertainty for not too large $M_{\rm SUSY}$, while 
the predictions using a pole mass matching are still reliable for $M_{\rm SUSY}$ of 1~TeV and even below. \\

\begin{figure}[hbt]
\centering
\includegraphics[width=0.75\linewidth]{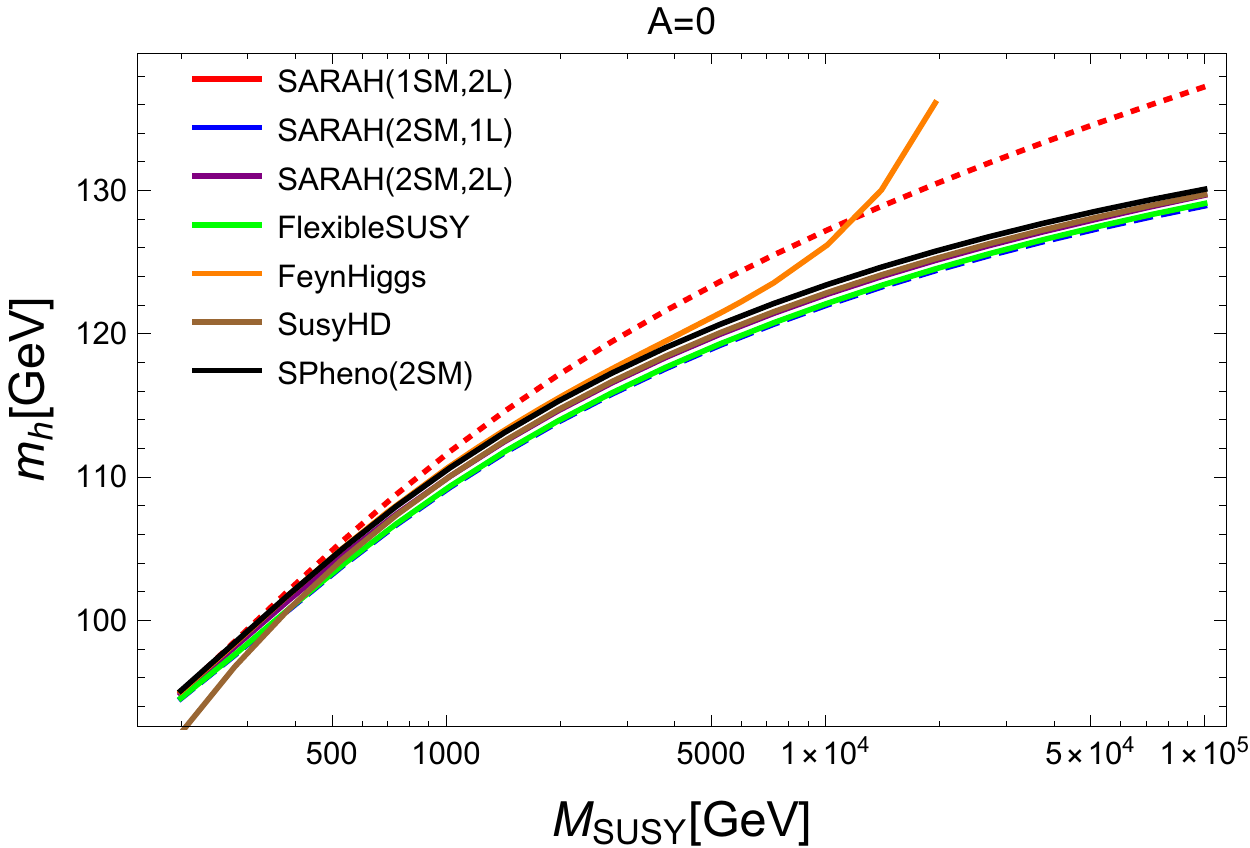} 
\caption{The Higgs mass prediction of different computer codes as function of the SUSY mass. The dashed red line corresponds to the old prediction
by \SARAH/\SPheno.}
\label{fig:MSUSY_mh}
\end{figure}

\begin{figure}[hbt]
\includegraphics[width=0.5\linewidth]{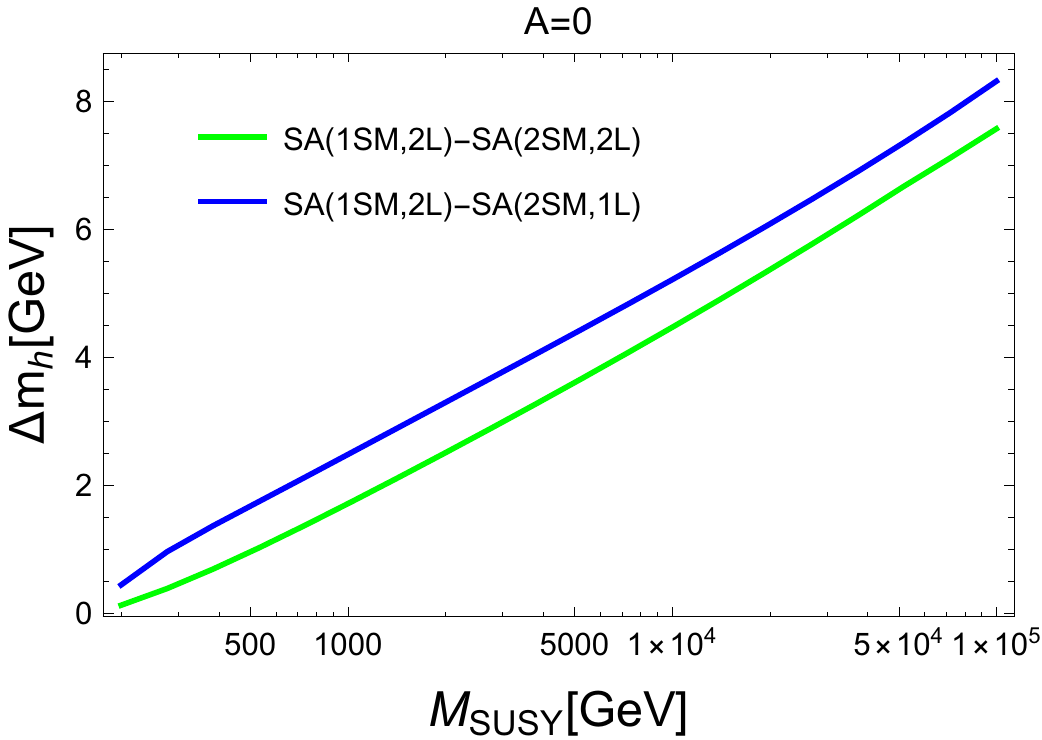} \hfill 
\includegraphics[width=0.5\linewidth]{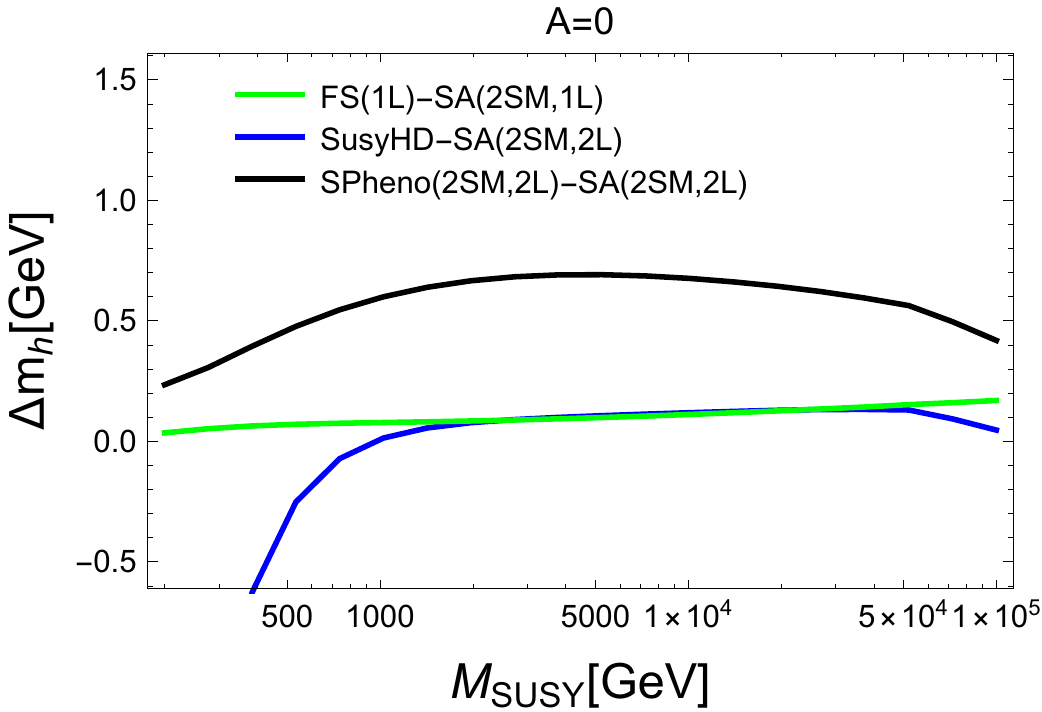} 
\caption{On the left: difference between the Higgs mass as predicted by the new \SARAH  and the old 
version using one- (blue) or two-loop (green) matching conditions for $\lambda_{\rm SM}$ at the SUSY scale. On the right: 
the differences between \SARAH and the new stand-alone \SPheno version (black), {\tt SusyHD} (blue; dashed line with three-loop 
thresholds to $Y_t$, full line without these corrections)  
as well as {\tt FlexibleSUSY}. We used here vanishing trilinear soft-breaking stop couplings.}
\label{fig:MSUSY_mh_diff}
\end{figure}

\begin{figure}[hbt]
\includegraphics[width=0.5\linewidth]{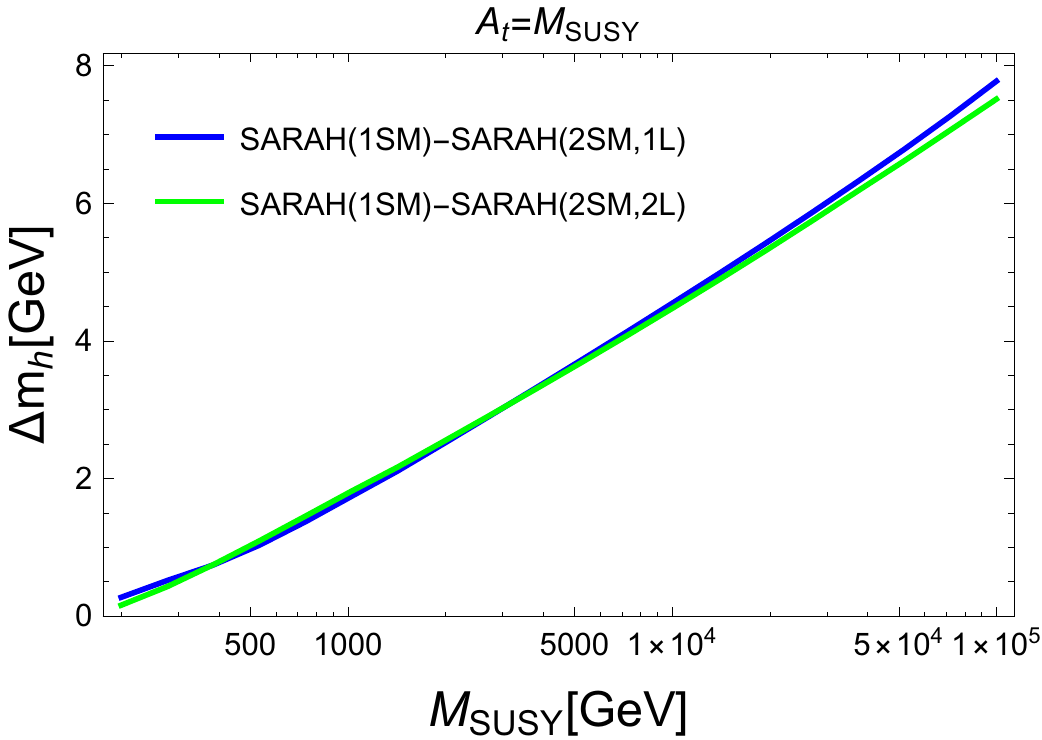} \hfill 
\includegraphics[width=0.5\linewidth]{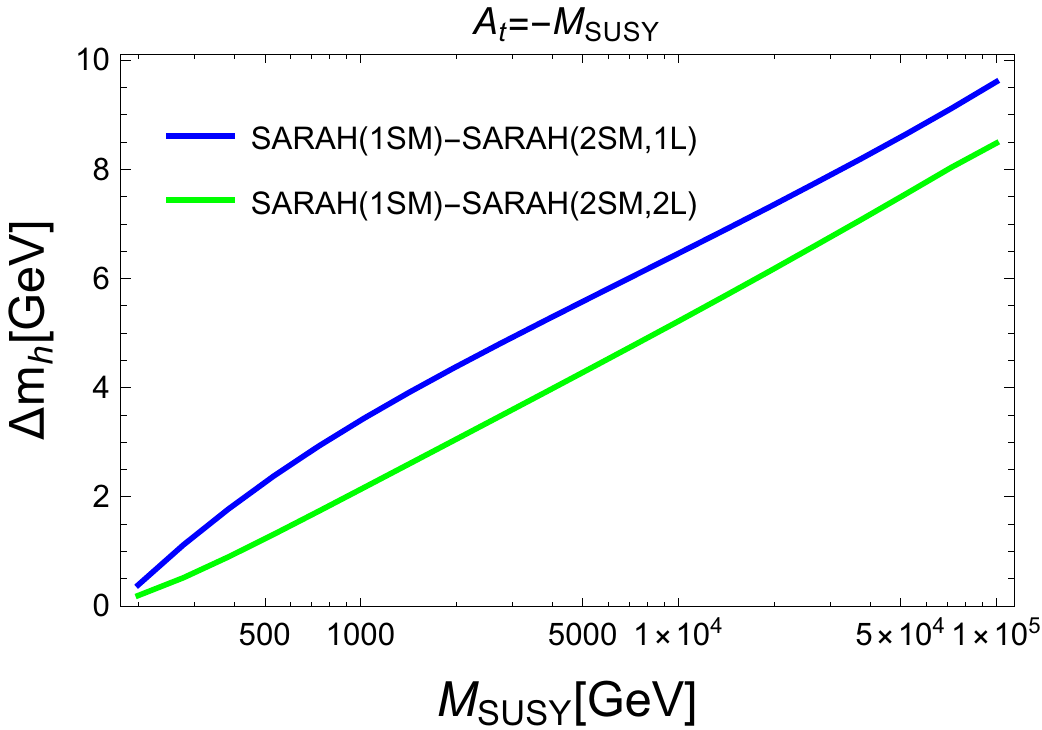} \\[2mm] 
\includegraphics[width=0.5\linewidth]{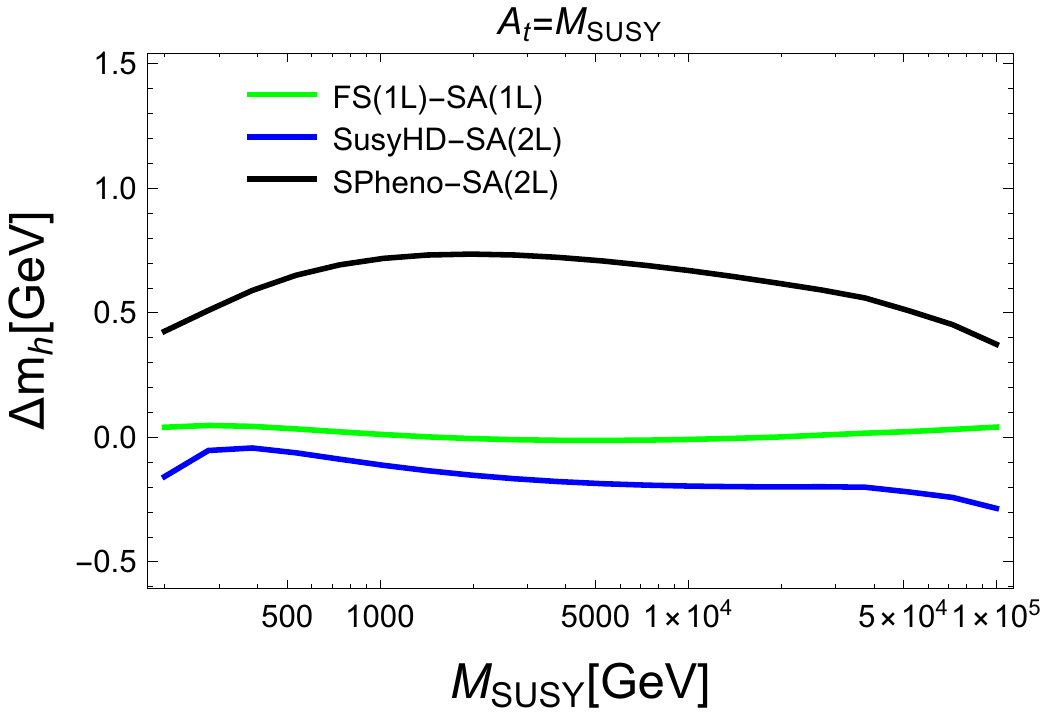} \hfill 
\includegraphics[width=0.5\linewidth]{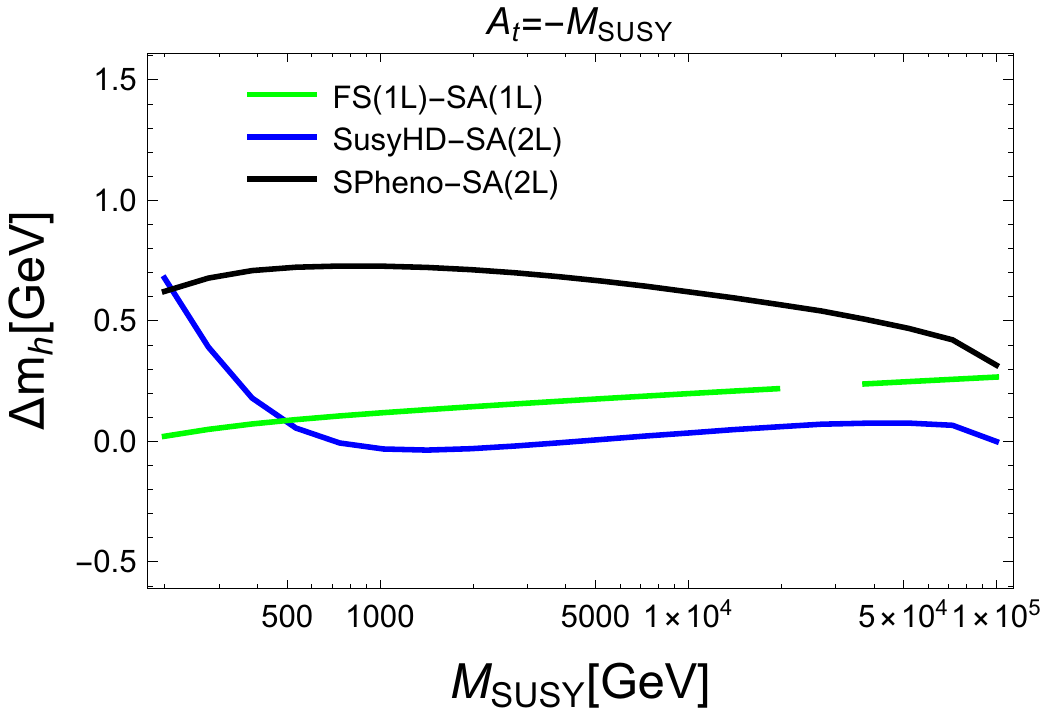} 
\caption{The same as Fig.~\ref{fig:MSUSY_mh_diff} for non-vanishing $A_t$. }
\label{fig:MSUSY_mh_diff_A0}
\end{figure}

We give a comparison of the Higgs mass prediction of the new \SARAH and \SPheno versions against
previous calculations as well as the current versions of {\tt FeynHiggs} (2.12.2), {\tt SusyHD} (1.0.2) 
and {\tt FlexibleSUSY} (1.7.2)\footnote{We used for the following comparison the model file {\tt MSSMtower} of {\tt FlexibleSUSY} 
which also performs a pole mass matching to get $\lambda_{\rm SM}(M_{\rm SUSY})$.}. 
For simplicity, we assume a degeneracy of the SUSY soft masses as well as 
$M_A$ and $\mu$ at the SUSY scale:
\begin{eqnarray}
&M_1 = M_2 = M_3 = M_A = \mu \equiv M_{\rm SUSY} & \\
&m_{\tilde e}^2 = m_{\tilde l}^2 = m_{\tilde d}^2 =m_{\tilde u}^2 =  m_{\tilde q}^2 = {\bf 1} M_{\rm SUSY}^2 &
\end{eqnarray}
We neglect all trilinear soft-terms but the one involving the stops which is parametrised as usual by
\begin{equation}
L = A_t Y_t \tilde{t}_L \tilde{t}^*_R H_u + \text{h.c} 
\end{equation}
The results for the Higgs mass prediction for $A_t = 0, \pm M_{\rm SUSY}$ and $M_{\rm SUSY}$ up to 100~TeV are summarised
in Figs.~\ref{fig:MSUSY_mh} -- \ref{fig:MSUSY_mh_diff_A0}. 
One can see in Fig.~\ref{fig:MSUSY_mh} that the new calculation of \SPheno/\SARAH gives a significant lower Higgs mass for very heavy 
SUSY scales and is in good agreement with the other codes like {\tt FlexibleSUSY} and {\tt SusyHD} for the entire 
range of $M_{\rm SUSY}$ shown here \footnote{The large rise in the Higgs mass 
as shown by {\tt FeynHiggs} for $M_{\rm SUSY} > 5$~TeV stems from a conversion problem of the 
input parameters and will most likely disappear in the near future \cite{Bahl}. 
}. {Only for small values of $M_{\rm SUSY}$ {\tt SusyHD} deviates from the other codes because of terms 
$O({v^2}/{M^2_{\rm SUSY}})$ missing due to the effective potential approach.} The main reason for the large rise  in the Higgs mass 
with \SPheno/\SARAH using a one-scale matching is the calculation of the top Yukawa coupling as discussed in sec.~\ref{sec:Coup}. 
Since the calculation is per se not wrong {but the differences in the calculation of $Y_t$ correspond 
to a three-loop effect in $m_h$}, the large changes in the Higgs mass prediction  shows how large the theoretical uncertainty 
of the fixed order calculation can become for very large SUSY scales. It might be surprising that a formal three-loop 
effect has such a big impact. However, it was for instance discussed in Ref.~\cite{Draper:2013oza} that at three-loop large cancellations appear, 
i.e. an incomplete three-loop calculation can give a quite misleading impression. 

Since the agreement between the different codes becomes impressively good even for very large SUSY masses, we give in Fig.~\ref{fig:MSUSY_mh_diff} the  
numerical differences between the Higgs mass predictions of \SARAH compared to the other codes. Also the difference between the one-scale matching and 
the two-scale matching using a one- or two-loop calculation of $\lambda$ is shown: for $M_{\rm SUSY} = 100$~TeV the Higgs mass prediction decreases 
by about 7~GeV when doing it via the EFT approach. The remaining differences to {\tt SusyHD} and {\tt FlexibleSUSY} is always better than 1~GeV, most often even better 
than 0.5~GeV \footnote{The public version of {\tt FlexibleSUSY} performs so far a one-loop matching for $\lambda$. We compare therefore the \SARAH results of a two-loop matching 
only with {\tt FeynHiggs}, \SPheno and {\tt SusyHD}, while we use for the comparison with {\tt FlexibleSUSY} the one-loop matching results.}. The increasing difference between \SARAH and {\tt FlexibleSUSY} compared
to \SPheno and {\tt SusyHD} comes from the calculation of the top Yukawa coupling in the SM: while \SARAH and {\tt FlexibleSUSY} use two-loop thresholds, \SPheno and {\tt SusyHD} have included even higher order corrections via the fit formula of Ref.~\cite{Buttazzo:2013uya}. These correction need not to be included because they are of a higher loop level than the Higgs mass calculation is done. Thus, the difference between these two calculations give an impression of the minimal, theoretical uncertainty which is at least present. The  differences between the codes also don't grow significantly if we use non-vanishing values for $A_t$ as shown in Fig.~\ref{fig:MSUSY_mh_diff_A0}: the overall changes in the Higgs mass between the \SARAH calculation in the full MSSM and in the effective SM changes again by 7--8~GeV for very large SUSY scales, while the difference to the other codes is in the range of 1~GeV and less.

\subsection{Higgs mass beyond the MSSM}
\begin{figure}[hbt]
\centering
\includegraphics[width=0.49\linewidth]{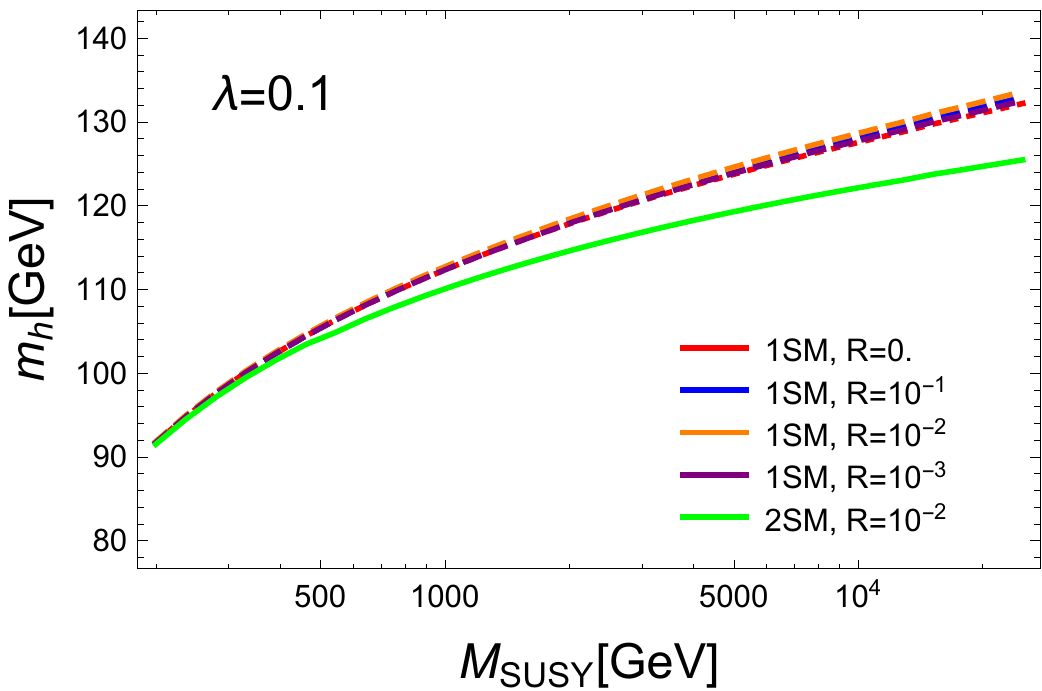} \hfill 
\includegraphics[width=0.49\linewidth]{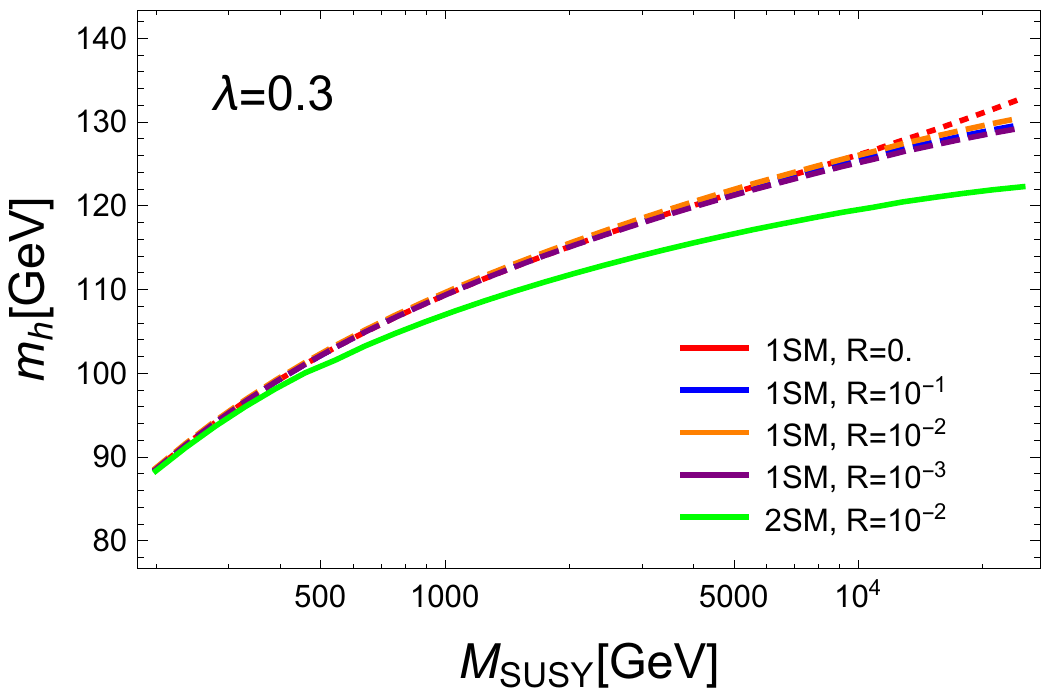} \\
\includegraphics[width=0.49\linewidth]{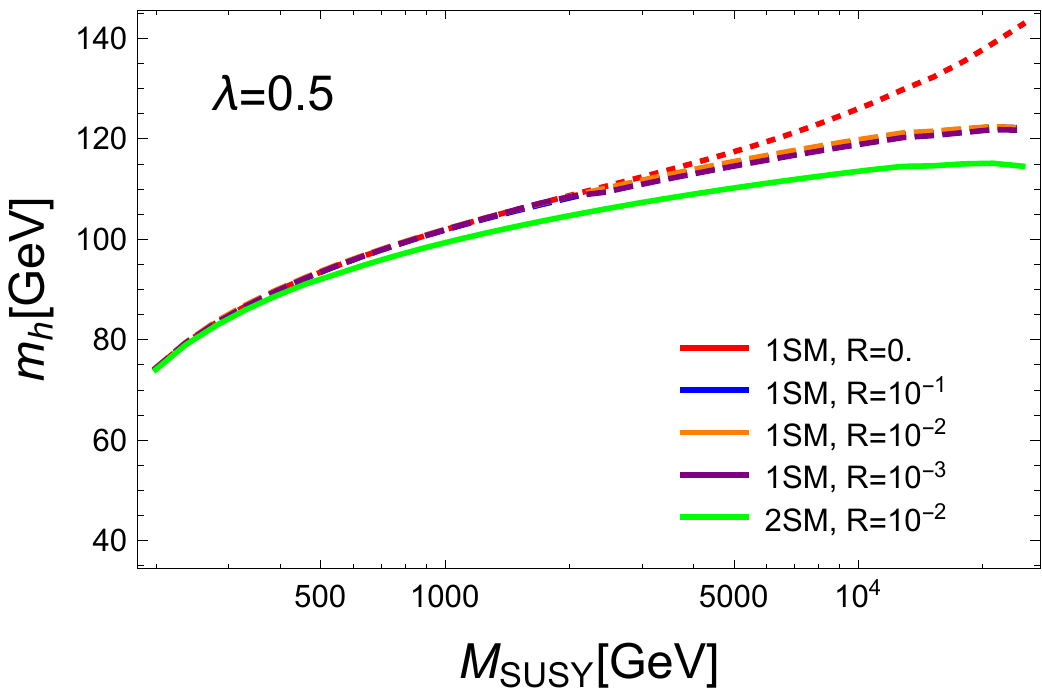} \hfill 
\includegraphics[width=0.49\linewidth]{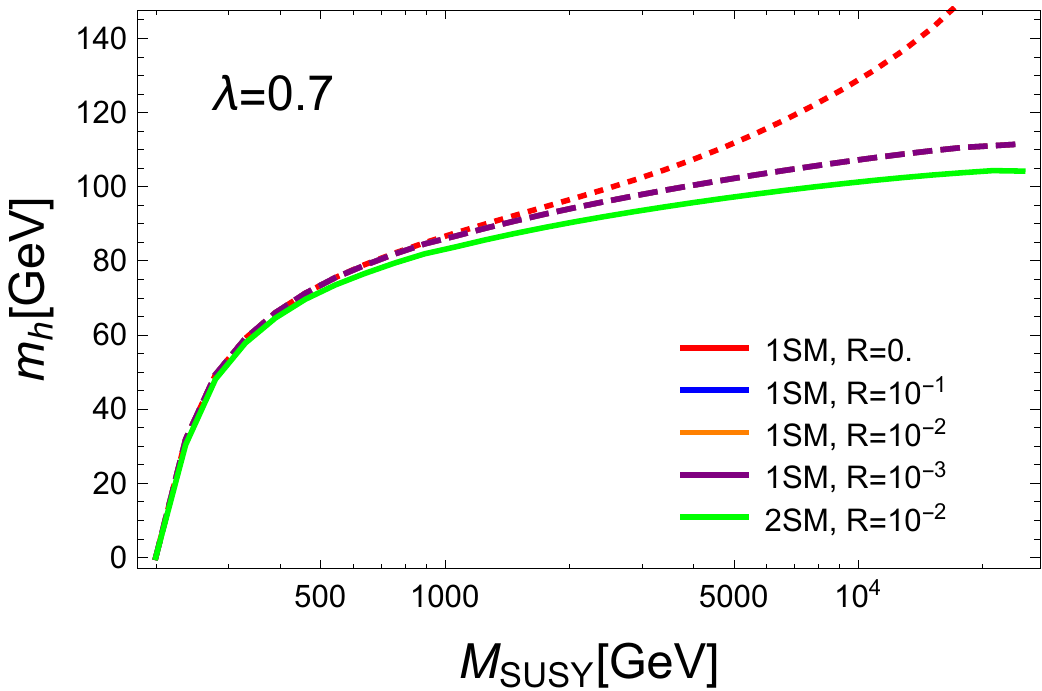} \\[4mm]
\includegraphics[width=0.49\linewidth]{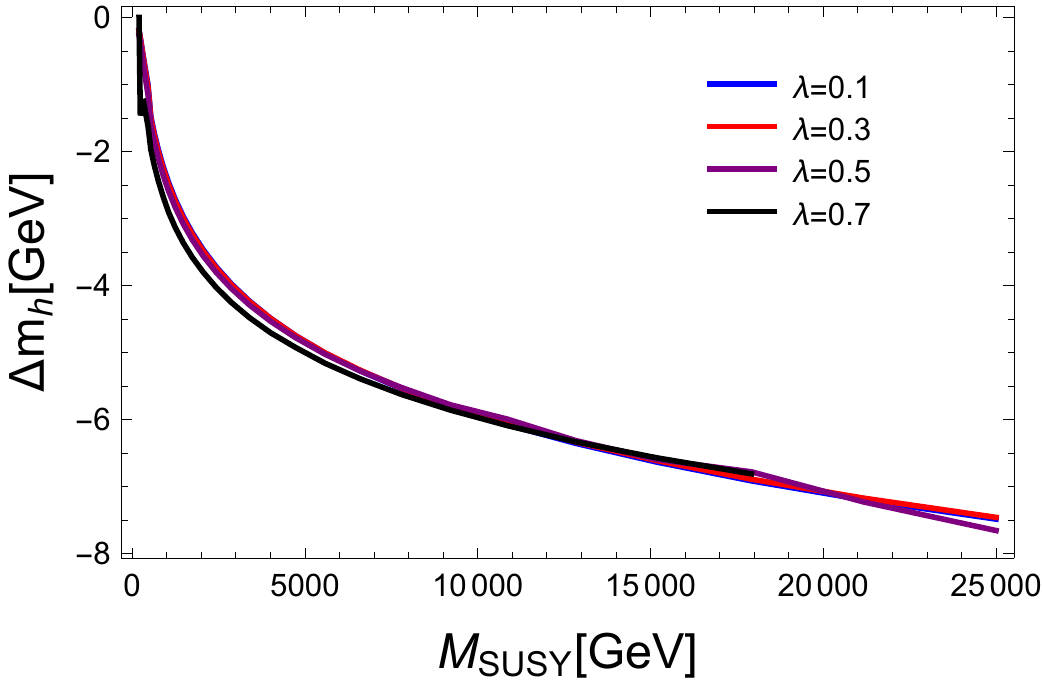} \\
\caption{The SM-like Higgs mass in the NMSSM as function of the SUSY scale for four different values of $\lambda$ (first, second row): the dotted red line gives 
the result of the one-scale matching (1SM) without regulator $R$, the dashed lines uses one-scale matching and $R=10^{-1},10^{-2},10^{-3}$, while for the green line the two-scale matching (2SM)
was used together with a Higgs mass calculation in the effective SM. The third row shows the difference $\Delta m_h$ between the one-scale and two-scale matching (both with $R=10^{-2}$).
}
\label{fig:NMSSM_mh}
\end{figure}

With \SARAH it is also possible to generate a spectrum generator for models beyond the MSSM which calculates mass spectra, 
decays and precision observables \cite{Porod:2014xia}. 
Also for these models 
two-loop Higgs mass calculations are performed by default. All important two-loop corrections stemming from new particles and/or new interactions 
are covered as discussed in detail in Refs.~\cite{Goodsell:2014bna,Goodsell:2015ira,Goodsell:2016udb}. The calculations make use of the generic results of
Refs.~\cite{Martin:2001vx,Martin:2002wn,Martin:2003it,Martin:2003qz,Martin:2005eg} and the only approximations used in the \SARAH implementation of the  
two-loop calculations are (i) the gaugeless limit, i.e. setting $g_1=g_2=0$, and (ii) neglecting momentum dependence, i.e.\ $p^2=0$. 
Thus, \SARAH provides for models beyond the MSSM the same precision in the Higgs mass as it does for the MSSM. 
Moreover,the obtained results with \SARAH include already for the next-to-minimal supersymmetric standard model (NMSSM) corrections which are not 
available otherwise \cite{Goodsell:2014pla,Staub:2015aea}. However, there is one additional subtlety when using these two-loop corrections in extended Higgs sector
which we need to discuss before coming to the results of the EFT approach: 
massless states appearing in the two-loop calculations usually cause divergences. Since the calculations are done in Landau gauge, these divergences are often 
associated with the Goldstone bosons of broken gauge groups what has caused the name 'Goldstone boson catastrophe' \cite{Martin:2013gka,Martin:2014bca}. For many cases this behaviour 
was already under control in \SARAH by the treatment of the $D$-terms what induced finite Goldstone masses as explained in Ref.~\cite{Goodsell:2016udb}.
However, for large SUSY scales, it can still happen that the ratio $m_S/M_{\rm SUSY}$ for some scalar mass $m_S$ becomes very small and introduces
numerical problems. As short-term workaround we have introduced for this reason a regulator $R$ which defines the minimal scalar mass squared as function 
of the renormalisation scale $Q$
\begin{equation}
m^2_{S,\rm min} = R Q^2 
\end{equation}
All scalar masses which appear in the two-loop integrals which are small than $m^2_{S,\rm min}$ are then replaced by $R Q^2$. 
We found that numerical dependence on $R$ is usually small for values of $R$ between 0.1 and 0.001. Nevertheless,  
the results of Ref.~\cite{Braathen:2016cqe} shall be included in \SARAH 
in the near future to have a rigorous solution to the Goldstone boson catastrophe which
is independent of any regulator \cite{workinprogress}. \\
We can turn now to the discussion of the changes in the Higgs mass prediction when using the EFT ansatz. In general, it 
is possible to use the two-scale matching together with an effective calculation of the Higgs mass within the SM also for non-minimal models. 
The procedure is exactly the same as for the MSSM. \SARAH uses the calculated Higgs mass in the full model to obtain $\lambda_{\rm SM}(M_{\rm SUSY})$ via 
a pole mass matching. It then evaluates  $\lambda_{\rm SM}(m_t)$ and calculates $m_h$ at that scale using SM corrections. 
We briefly discuss the impact of the new calculation at the example of the NMSSM\footnote{We refer to Ref.~\cite{Ellwanger:2009dp} for an introduction 
into the NMSSM and for questions regarding the notation in the following}. For this purpose, we relate the NMSSM specific, 
dimensionful parameters to the SUSY scale via
\begin{eqnarray*}
&\mu_{\rm eff} = M_{\rm SUSY}\,,\hspace{1cm} A_\kappa = -\lambda M_{\rm SUSY}\,,
\hspace{1cm} A_\lambda= M_{\rm SUSY}\left(\frac{\tan\beta}{(1+\tan\beta^2)} - \frac{\kappa}{\lambda}\right)&
\end{eqnarray*}
With this parametrisation we find that the heavy MSSM-like scalars get a tree-level mass of $M_{\rm SUSY}$ while also the scalar singlets are sufficiently heavy to be integrated out at $M_{\rm SUSY}$. We set in addition 
\begin{eqnarray*}
 \tan\beta = 4 \,,\hspace{1cm} \lambda = \kappa
\end{eqnarray*}
Thus, the only free parameters left are $\lambda$ and $M_{\rm SUSY}$.  The Higgs mass for a variation of $M_{\rm SUSY}$ for $\lambda=0.1,0.3,0.5,0.7$ is shown in Fig.~\ref{fig:NMSSM_mh}. Here, we also show the 
results with and without regulator $R$. One can see that the numerical problems associated with small masses, which in this case here are the light Higgs as well as the 
two Goldstone bosons, show up for increasing $M_{\rm SUSY}$. The larger 
$\lambda$ is, the more pronounced these problems are. However, with a regulator $R=0.01$ this behaviour can be prevented 
for all values of $\lambda$ and $M_{\rm SUSY}$ shown here for the one- and two-scale matching.  We find that the results with regulator masses 
are in agreement with Ref.~\cite{Athron:2016fuq} within the indicated uncertainties.  \\
The impact on the Higgs mass using the new two-scale matching is similar as for the MSSM: for SUSY masses up to 2~TeV, the effects are small and 
less than 2~GeV, but they quickly increase with increasing $M_{\rm SUSY}$. For $M_{\rm SUSY} = 25$~TeV, the difference in the Higgs mass prediction
is between 5.5 and 6.5~GeV. For our example we find that the differences depend only weakly on the value of $\lambda$. 

Similarly, one can use now \SARAH to study also the Higgs masses for other models in the presence of large SUSY scales more precisely. However, a detailed 
exploration of these effects in other models is beyond the scope of this paper. Here, we want to stress
that one should be careful with models with extended Higgs sector because not all scalar masses become automatically large if $M_{\rm SUSY}$ is large. Examples are 
for instance models with extended gauge sectors in which a second light scalar can appear because of $D$-flat directions \cite{OLeary:2011vlq,Hirsch:2011hg,Hirsch:2012kv}.
In these cases, a sizeable mixing 
between the SM-like Higgs and another scalar can be present, i.e. the calculation of $m_h$ within an effective SM might now be valid. Therefore, \SARAH does not perform
this calculation by default, if a second CP-even scalar with a mass below 500~GeV is present.  

\subsection{Perturbativity limit of new interactions}
Many models beyond the MSSM are attractive because they give a tree-level enhancement of the Higgs mass. 
This is quite interesting from the point of view because it reduces the required loop contributions to obtain
$m_h=125.1$. Usually this allows for smaller values of $A_t$ which is important for the stability
of the scalar potential \cite{Camargo-Molina:2013sta,Blinov:2013fta,Chowdhury:2013dka,Camargo-Molina:2014pwa,Beuria:2016cdk}. 
The best studied example is again the NMSSM which pushes the Higgs mass via new $F$-term contributions which are proportional to $\lambda^2$. 
We demonstrate this in Fig.~\ref{fig:NMSSM_lam_max} where we compare the dependence of the Higgs mass on the stop mixing parameter $X_t$ as defined 
as
\begin{equation}
X_t = A_t -\mu\tan\beta 
\end{equation}
In the NMSSM, $\mu$ is replaced by $\mu_{\rm eff}$. 
We see for a SUSY scale of 5~TeV and the  chosen value of $\tan\beta=2$ and $\lambda=0.6$ even without stop mixing the Higgs mass can be found in the correct mass range of 122-128~GeV.

\begin{figure}[hbt]
\centering
\includegraphics[width=0.49\linewidth]{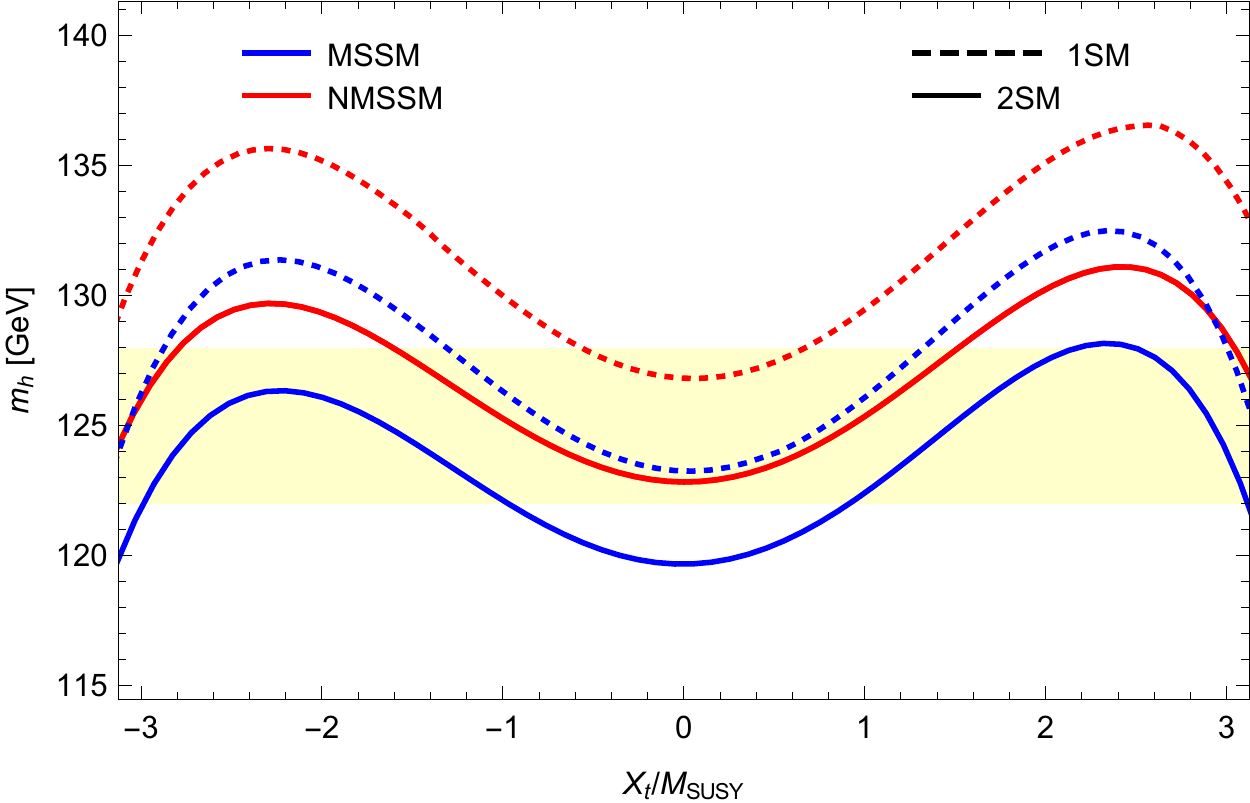}  
\caption{The Higgs mass in the MSSM and NMSSM as function of $X_t/M_{\rm SUSY}$ using one- and two-scale matching. 
Here we set $\mu = M_{\rm SUSY} = 5$~TeV and used for the MSSM $\tan\beta=10$, $M_A=5$~TeV. The input parameters for the NMSSM were $\lambda=0.6$, $\kappa=0.2$,
$A_\lambda=10$~TeV, $A_\kappa=-5$~TeV, $\tan\beta=2$.}
\label{fig:NMSSM_lam_max}
\end{figure}

Because of this large impact of $\lambda$ on the Higgs mass , it is very important to know how big $\lambda$ can be in order to be still in agreement with gauge couplings unification at $M_{GUT}$. 
\begin{figure}[hbt]
\centering
\includegraphics[width=0.49\linewidth]{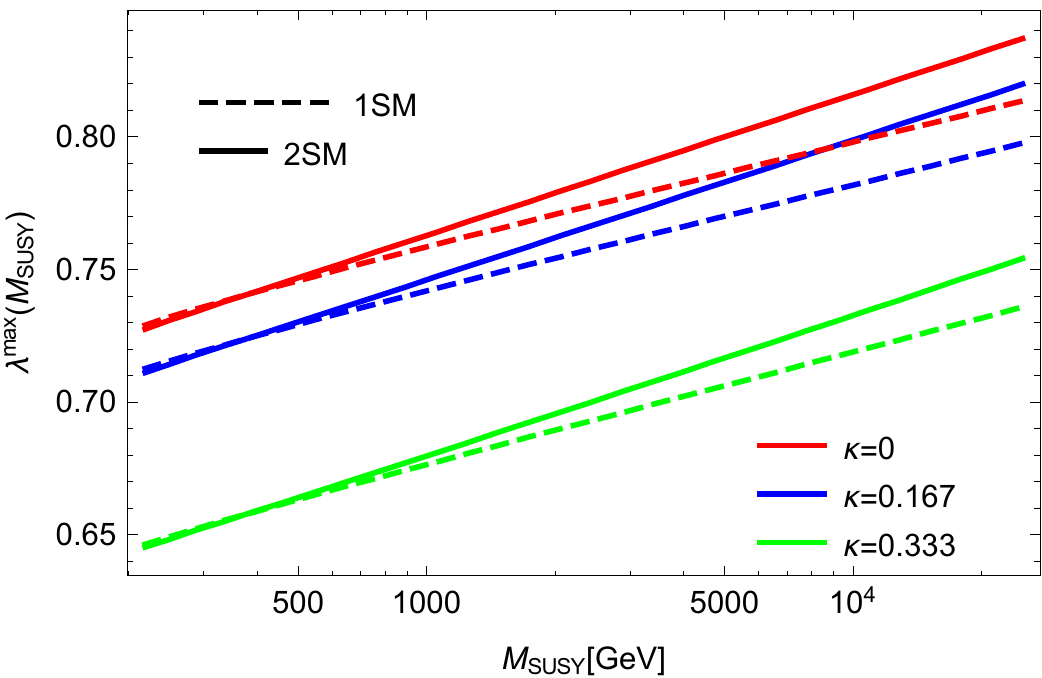} 
\includegraphics[width=0.49\linewidth]{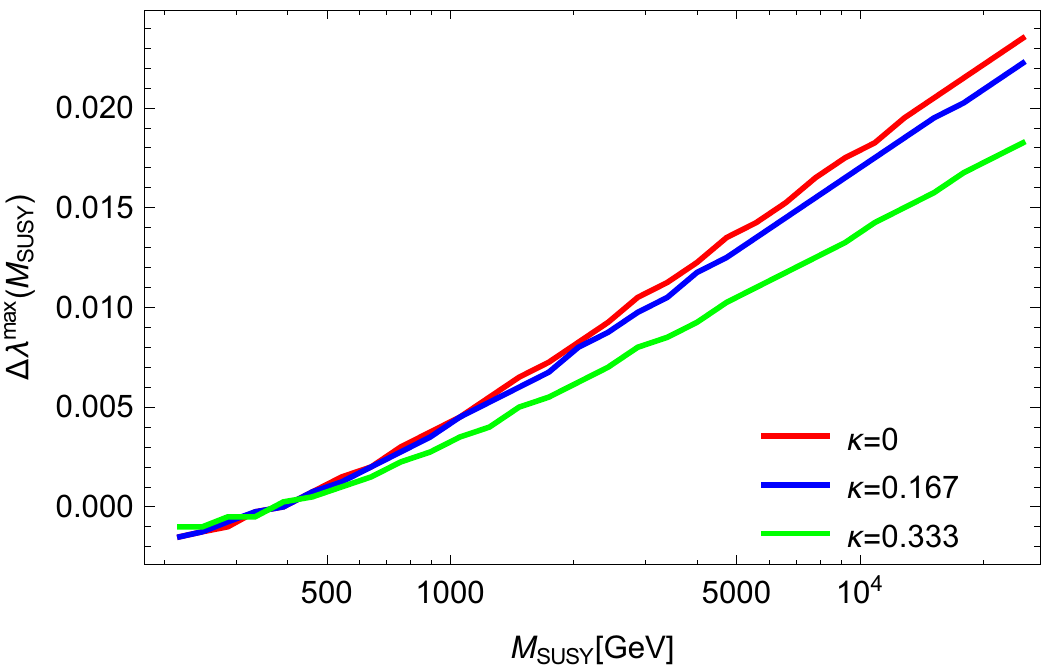} 
\caption{Left: maximal value of $\lambda(M_{SUSY}$ consistent with perturbativity up to $M_{GUT}$
for different values of $\kappa$. The full (dashed) lines correspond to the case of
two (one) scale matching. Right: the difference  $\Delta\lambda = \lambda^{\rm max}_{\rm 2SM}(M_{\rm SUSY})-\lambda^{\rm max}_{\rm 1SM}(M_{\rm SUSY})$ of the two 
matching schemes.}
\label{fig:NMSSM_lam_max}
\end{figure}

In Figure~\ref{fig:NMSSM_lam_max} we display the maximal value of $\lambda(M_{\rm SUSY})$ which does not lead to a 
Landau pole below $M_{GUT}$ for different values of $\kappa(M_{\rm SUSY})$ 
and for  $M_{\rm SUSY}$ up to 25~TeV and $\tan\beta=4$, and show the differences between the 
one- and two-scale matching. Because of the smaller top Yukawa coupling in the two-scale approach, one finds that slightly larger values of $\lambda(M_{\rm SUSY}$ 
are allowed that for the one-scale matching.

\section{Conclusion}
\label{sec:conclusion}
We have presented the new two-scale matching procedure in \SARAH/\SPheno to improve the prediction of the running \DR gauge and Yukawa couplings 
at the SUSY scale for large values of $M_{\rm SUSY}$. Together with the new matching, also the possibility of an EFT Higgs mass calculation is 
introduced. In the EFT calculation $\lambda_{SM}$ is obtained via a Higgs pole mass matching at $M_{\rm SUSY}$ and the SM-like Higgs mass is calculated
within the SM at the top mass scale. We have shown various consequences of the two-scale matching and the EFT Higgs mass calculation in the MSSM and 
beyond. In particular, we have compared the Higgs mass prediction for SUSY scales up to 100~TeV and found a good agreement with other EFT codes 
as {\tt SusyHD} and {\tt FlexibleSUSY}. We have also shown that the value of $\mu$ in the CMSSM can change significantly because of the changes in the 
top Yukawa coupling. This has an direct impact on naturalness considerations.

\section*{Acknowledgements}
We thank Alexander Voigt for helpful discussions concerning the matching procedure in {\tt FlexibleSUSY} and
Eliel Camargo for his contribution in the early stage of this work. W.P.~has been supported
by the DFG,  project nr.~PO 1337/7-1.

%% file: appendix.tex
\section{Matching}
\label{app:Matching}

\subsection{One scale matching}
\label{app:OneScale}
Before we present the new two-scale matching which is now performed by \SARAH/\SPheno, we review the 
current procedure. The first step is that all \DR parameters are calculated already at $\MZ$ and 
two-loop SUSY RGEs are used for the running to $M_{\rm SUSY}$.

\subsubsection{Strong coupling}
The strong interaction coupling at the weak scale is matched to the input value 
$\alpha^{(5)}_s(\MZ)$ in the $N_f=5$ flavour scheme via
\begin{align}    \alpha_s^{\DR}(\MZ) &= \frac{\alpha_s^{(5),\overline{\text{MS}}}(\MZ)}{1 - \Delta\alpha_s(\MZ)},  \\
    \Delta\alpha_s(\MZ) =& \frac{\alpha_s}{2\pi} \left( \frac{1}{2} - \frac{2}{3}
      \log{\frac{m_t}{\MZ}} + \Delta_s^{\rm MSSM} \right),
\end{align}
The corrections due to the new coloured states in the MSSM are given by
\begin{equation}
\Delta_s^{\rm MSSM} = -
      2 \log{\frac{m_{\tilde g}}{\MZ}} - \frac{1}{6} \sum_{i=1}^6 \left(
        \log{\frac{m_{\tilde u_i}}{\MZ}} + \log{\frac{m_{\tilde
              d_i}}{\MZ}}\right)
\end{equation}
For any other BSM model, $\Delta_s^{\rm MSSM}$ is adjusted by \SARAH to fit to the particle content.

\subsubsection{Electroweak sector} 
The EW gauge sector of the MSSM is determined by four fundamental parameters. 
These are usually the gauge couplings for $SU(2)_L \times U(1)_Y$ and 
the electroweak VEVs for the up- and down-Higgs
\begin{eqnarray*}
g_1\,,\hspace{0.5cm}  g_2\,,\hspace{0.5cm} v_d\,,\hspace{0.5cm} v_u
\end{eqnarray*}
$v_d$ and $v_u$ are derived from the calculated ew VEV $v(\MZ)^2 = \sqrt{v_d^2+v_u^2}$ and 
the input value for $\tan\beta=\frac{v_u}{v_d}$ which could either be given at $\MZ$ or 
$M_{\rm SUSY}$. Thus, the matching procedure needs to determinate $v^{\DR}(\MZ)$, $g^{\DR}_1(\MZ)$ and
$g^{\DR}_2(\MZ)$ from three physical quantities. Here, \SPheno and \SARAH use as input the $Z$ mass, the Fermi constant $G_F$ and the electromagnetic coupling of the SM at the
scale $\MZ$ in the 5-flavour scheme, $\alpha_{em}^{(5),\overline{\text{MS}}}(\MZ)$. \\

The relations between the input and \DR parameters is as follows:
\begin{enumerate}
 \item The electroweak coupling constant is calculated from 
 
 \begin{align}
  \alpha_{em}^{\DR}(\MZ) &= \frac{\alpha_{em}^{(5),\overline{\text{MS}}}(\MZ)}{1 - \Delta\alpha(\MZ)} ,\\
    \Delta\alpha(\MZ) =& \frac{\alpha}{2\pi} \Big(\frac{1}{3}-
      \frac{16}{9}  \log{\frac{m_{t}}{\MZ}} + \Delta_{em}^{\rm MSSM} \Big) .
\end{align}
with 
\begin{align}
\Delta_{em}^{\rm MSSM} = & - \frac{4}{9} \sum_{i=1}^6  \log{\frac{m_{\tilde u_i}}{\MZ}} - \frac{1}{9} \sum_{i=1}^6
      \log{\frac{m_{\tilde d_i}}{\MZ}}  \nonumber \\
      & \hspace{1cm}  -\frac{4}{3} \sum_{i=1}^2 \log{\frac{m_{\tilde \chi^+_i}}{\MZ}} -  \frac{1}{3} \sum_{i=1}^6 \log \frac{m_{\tilde{e}_i}}{\MZ} - \frac{1}{3}  \log{\frac{m_{H^+}}{\MZ}}
\label{eq:delta_alpha}      
\end{align}
Again, if another model shall be considered, the value of $\Delta_{em}$ is calculated by \SARAH automatically.

\item The Weinberg angle $\sin^{\DR}\Theta_W$ at the scale $\MZ$ is obtained iteratively from the above-computed $ \alpha_{em}^{\DR}(\MZ)$,
    together with $G_F$ and $\MZ$,  via
\begin{align}
  \left(\sin^{\DR}\Theta_W \cos^{\DR}\Theta_W\right)^2 =
  \frac{\pi\,\alpha_{em}^{\DR}(\MZ)}{\sqrt{2} \MZ^2 G_F (1-\delta_r)} ,
\end{align}
where we have introduced
\begin{align}
\label{def:deltar}
  \delta_r &= \hat\rho \frac{\Pi_{WW}^T(0)}{M_W^2} -  \frac{\re\Pi_{ZZ}^T(\MZ^2)}{\MZ^2} + \delta_{\rm VB} + \delta_r^{(2)} , \\
  \hat\rho &= \frac{1}{1-\Delta\hat\rho} ,\qquad\qquad \Delta\hat\rho
  = \re\Biggl[ \frac{\Pi_{ZZ}^T(\MZ^2)}{\hat\rho\,\MZ^2} - \frac{\Pi_{WW}^T(M_W^2)}{M_W^2}\Biggr] + \Delta\hat\rho^{(2)}~,
\end{align}
Here, $\Pi^T_{VV}(p^2)~~(V = Z,W)$ are the \DR-renormalized transverse parts of the self-energies of the vector bosons, computed at the
renormalization scale $Q=\MZ$, and $\delta_r^{(2)}$ and $\Delta\hat\rho^{(2)}$ are two-loop corrections as given in
\cite{Fanchiotti:1992tu,Pierce:1996zz}  
\begin{align}
\delta_r^{(2)} =  \frac{f_1}{(1-\sin^2\Theta^{\MS}_W ) \sin^2\Theta^{\MS}_W} - x_t (1 - \delta_r) \rho 
\end{align}
with
\begin{align}
x_t = &  3 \left(\frac{G_F m_t^2}{8 \pi^2}\right)^2 \rho_2\left(\frac{m_h}{m_t}\right) \\
f_1 = & \frac{\alpha_S^{\MS} \alpha_{ew}^{\MS}}{4\pi^2} \left(2.145 \frac{m_t^2}{\MZ^2}+0.575 \log\frac{m_t}{\MZ}-0.224-0.144 \frac{\MZ^2}{m_t^2}\right) \\
f_2 = & \frac{\alpha_S^{\MS} \alpha_{ew}^{\MS}}{4\pi^2} \left(-2.145 \frac{m_t^2}{\MZ^2}+1.262 \log\frac{m_t}{\MZ}-2.24-0.85 \frac{\MZ^2}{m_t^2}\right) 
\end{align}
and 
\begin{align}
\rho_2(r)= & 19-16.5r+\frac{43}{12} r^2+\frac{7}{120} r^3-\pi \sqrt{r} (4-1.5 r+\frac{3}{32} r^2 +\frac{r^3}{256}) \nonumber \\ 
& -\pi^2 (2-2 r+0.5 r^2)-\log(r) (3 r-0.5 r^2) 
\end{align}
The one-loop vertex and box corrections $\delta_{\rm VB}$ implemented into \SPheno are hard-coded and taken from literature\cite{Degrassi:1990tu, Grifols:1984xs, Chankowski:1993eu}, while the ones used by \SARAH are auto-generated and include therefore all one-loop corrections beyond the MSSM. Also the self-energies $\Pi^T$ are automatically calculated by \SARAH 
at the full one-loop level. 
\item The electroweak VEV $v$ used to calculate $v_d$ and $v_u$ at $\MZ$ is obtained from
\begin{equation}
v^{\DR}(\MZ)= \sqrt{{\MZ^{\DR}(\MZ)^2}~\frac{(1- \sin^2\Theta^{\DR}_W)\sin^2\Theta^{\DR}_W}{\pi \alpha^{\DR}} }.
\end{equation}
Here, the running mass $\MZ^{\DR}$ is given by
\begin{align}
M^{\DR}_Z(M)Z) = \sqrt{\MZ^2 + \Pi^T_{ZZ}(\MZ^2)}
\end{align}
%
\end{enumerate}

\subsubsection{Yukawa couplings}
In order to calculate the value of the \DR-renormalized Yukawa coupling at the SUSY scale,  
\SPheno used so far the approach of  Ref.~\cite{Pierce:1996zz}.
First, for all leptons and the five light quarks the \DR masses at $\MZ$ are calculated. 
Afterwards, the additional non-SUSY thresholds stemming from massive bosons and the full  one-loop SUSY 
thresholds are included. For $m_t$ 
also the known two-loop QCD  corrections are added \cite{Avdeev:1997sz,Bednyakov:2002sf}
\begin{equation}
\Sigma^{(2)}_{\rm QCD} = \frac{1}{16\pi^2}{\alpha_s}{18} \left(2011 - 1476 \log(Q) + 396 (\log(Q))^2  - 48 \zeta_3 + 
   16 \pi^2 (1 + 2 \log2)\right)
\end{equation}

Using these loop corrections, the loop-corrected $3 \times 3$ mass matrices for quarks and leptons are 
calculated via
\begin{equation}
\label{eq:oneloopMF}
m_f^{(1L)}(p^2_i) =  m_f^{(T)} - {\Sigma}_{S,f}(p^2_i) - {\Sigma}_{R,f}(p^2_i) m_f^{(T)} - m_f^{(T)} {\Sigma}_{L,f}(p^2_i) 
\end{equation}
with $f={l,d,u}$. Here, $\Sigma_{S,R,L}$ are usually the one-loop self-energies {\it without} photon and gluon corrections. 
Only for the top-quark photon and gluon corrections need to be included and in addition one identifies
\begin{equation}
{\Sigma}_{S,t} =  \Sigma^{(1)}_{S,t} + \Sigma^{(2)}_{QCD}
\end{equation}
The \DR Yukawa matrices fulfilling 
\begin{equation}
m_u^{(T)} = \frac{1}{\sqrt{2}} Y_u v_u \,, \hspace{1cm} m_d^{(T)} = \frac{1}{\sqrt{2}} Y_d v_d \,,  \hspace{1cm} m_l^{(T)} = \frac{1}{\sqrt{2}} Y_l v_l \,,
\end{equation}
are calculated iteratively from eq.~(\ref{eq:oneloopMF}) by the condition that the eigenvalues of 
$m_f^{(1L)}(p^2_i)$ must coincide with the \DR values for the light leptons and the top pole mass respectively.

\subsection{Two scale matching}
\label{app:TwoScale}
In the new two scale approach, the separation of the matching is that all SM corrections are included at $\MZ$ to obtain the \MS values which are then shifted at $M_{\rm SUSY}$ to 
their \DR values by including all one-loop SUSY thresholds. 

\subsubsection{Calculating the \MS parameters at $\MZ$}
The calculation of the \MS parameters at $\MZ$ is very similar to the approach described in the last section, but with all BSM 
contributions removed. 
\begin{enumerate}
\item We get for the gauge couplings
\begin{align}
 \alpha_S^{\MS} = &  \frac{\alpha_s^{(5),\overline{\text{MS}}}(\MZ)}{1 + \frac{2}{3} \frac{\alpha_s}{2\pi} \left(  \log{\frac{m_t}{\MZ}} \right)}  \\
 \alpha_{ew}^{\MS} = &   \frac{\alpha_{em}^{(5),\overline{\text{MS}}}(\MZ)}{1 + \frac{\alpha}{2\pi} \Big(  \frac{16}{9}  \log{\frac{m_{t}}{\MZ}} \Big) } \\
\end{align}
\item The Weinberg angle is calculated as
\begin{align}
\sin\Theta^{\MS}_W = & \frac{1}{2} - \sqrt{\frac{1}{4}-\frac{\pi \alpha_{ew}^{\MS}(\MZ)} {\sqrt{2} \MZ^2 G_F (1 -\delta_r)} } 
\end{align}
with  $\delta_r$ defined in eq.~(\ref{def:deltar}). The following one-loop SM contributions are used:
\begin{align}
\delta_{VB} = g_2^{\MS,2} \rho \left(6 + \frac{\log \cos^2\Theta_W}{\sin^2\Theta_W} \left(\frac{7}{2} - \frac{5}{2} \sin^2\Theta_W - \sin^2\Theta^{\MS}_W  \left(5 + \frac{3}{2} \frac{\cos^2\Theta_W}{\cos^2\Theta^{\MS}_W} \right) \right)\right)  
\end{align}
and the two-loop corrections $\delta_r^{(2)}$ agree with the ones used in the one scale matching. 
\item The VEV is obtained from
\begin{equation}
v^{\MS} =  (\MZ^{2,\MS}(M_{\rm \MZ})+\delta \MZ^{2,\MS}) \frac{(1-\sin\Theta^{\MS}_W) \sin\Theta^{\MS}_W}{\pi  \alpha^{\MS}_{ew}(M_{\rm \MZ}) } 
\end{equation}
where $\delta \MZ = \Pi^T_{ZZ}(\MZ^2)$ includes only the SM corrections.  
\item The Yukawa couplings are obtained from the running $\MS$ quark and lepton masses. Here, we include for $m_t$ the two-loop corrections to relate the \MS and pole mass\cite{Fleischer:1998dw}
\begin{align}
m_t^{\MS} = & m_t^{\rm pole} \Big[1+ \frac{1}{16\pi^2} \left( \left(\frac{16 \pi}{9}\alpha - \frac{16 \pi}{3} \alpha_s \right) (4 + \log(Q))  \right) \nonumber \\
& - \frac{1}{(16\pi^2)} \frac{\alpha_s^2}{18} \left(2821 + 2028 \log(Q) + 396  (\log(Q))^2 + 16 \pi^2 (1+ 2 \log2) - 48 \zeta_3 \right) \Big] 
\end{align}
The \MS Yukawa matrices are calculated iteratively from the condition that the \MS fermion masses are reproduced once the one-loop SM corrections with massive bosons 
are included: 
\begin{equation}
\label{eq:oneloopMF}
m_f^{(1L)}(p^2_i) =  m_f^{(T)} - \tilde{\Sigma}_S(p^2_i) - \tilde{\Sigma}_R(p^2_i) m_f^{(T)} - m_f^{(T)} \tilde{\Sigma}_L(p^2_i) 
\end{equation}
Here, $\tilde{\Sigma}$ are the self-energies without the photonic and gluonic contributions. 
The eigenvalues of $m_f^{(1L)}(p^2_i)$ must coincide with $m_{f}^{\MS}(\MZ)$.
\end{enumerate}

$g_i^{\MS}$ ($i=1,2,3$), $Y_f^{\MS}$ ($f=l,d,u$) and $v^{\MS}$ are then evaluated from $\MZ$ to $M_{SUSY}$ using the full two-loop SM RGEs
which are extended by the three-loop contributions involving $g_3$, $\lambda$ and $Y_t$. \\

For the top Yukawa and strong gauge coupling one can include in \SPheno an additional threshold at $m_t$ at which higher order corrections are included by using the fit formulae \cite{Buttazzo:2013uya}
\begin{align}
Y_t(m_t) = & 0.9369 + 0.00556 \left(\frac{m_t}{\rm GeV} - 173.34\right) - 0.6 (\alpha_s(\MZ)-0.1184) \\
g_3(m_t) = & 1.1666 + 0.00314 \frac{(\alpha_s(\MZ)-0.1184)}{0.0007} - 0.00046 \left(\frac{m_t}{\rm GeV} - 173.34\right) 
\end{align}

\subsubsection{Calculating the \DR parameters at $M_{SUSY}$ in \SARAH}
At the $M_{SUSY}$, the \MS parameters are first shifted to \DR parameters and then the SUSY thresholds are added.
\begin{enumerate}
 \item {\bf Strong coupling}
\begin{align}
  \alpha^{\DR}_{S}(M_{\rm SUSY}) =  \frac{\alpha^{\MS}_{S}(M_{\rm SUSY})}{1-\Delta^{\DR}_{\alpha_S}}
\end{align}
with 
\begin{equation}
\Delta^{\DR}_{\alpha_S} =\frac{\alpha_s}{2\pi} \left( \frac{1}{2} -  \Delta_s^{\rm MSSM}  \right)
\end{equation}

 \item {\bf Electroweak sector}: \\
The electroweak gauge coupling is calculated from $g_1^{\MS}$, $g_2^{\MS}$ and translated into its  $\DR$  value via
\begin{align}
 \alpha^{\MS}_{ew}(M_{\rm SUSY}) = & \frac{1}{4\pi} \frac{(g_1^{\MS} g_2^{\MS})^2}{(g_1^{\MS})^2+(g_2^{\MS})^2} \\
  \alpha^{\DR}_{ew}(M_{\rm SUSY}) =  \frac{\alpha^{\MS}_{ew}(M_{\rm SUSY})}{1-\Delta^{\DR}}
\end{align}
with 
\begin{equation}
\Delta^{\DR} = \frac{\alpha^{\DR}_{ew}}{2\pi} \left(\frac{1}{3} + \Delta_{em}^{\rm MSSM} \right)
\end{equation}
where $\MZ$ has to be replace by $M_{\rm SUSY}$ in eq.~(\ref{eq:delta_alpha}).
In addition, it is helpful to define for later use
\begin{align}
\sin\Theta^{\MS}_W =& \frac{g_1^{\MS}}{\sqrt{(g_1^{\MS})^2+(g_2^{\MS})^2)}}  \\
\delta_r^{\MS} =& 1 - \frac{\pi \alpha^{\MS}_{ew}(M_{\rm SUSY})}{\sqrt{2} G_F \MZ^2  \sin^2\Theta^{\MS}_W (1- \sin^2\Theta^{\MS}_W)}
\end{align}
as well as
\begin{align}
\delta_{VB}^{\DR} &=  \delta_{VB}^{\rm MSSM} - \delta_{VB}^{\rm SM} \\
\delta \MZ^{2,\DR} &= \Pi_{ZZ}^{T,\rm MSSM} - \Pi_{ZZ}^{T,\rm SM} \\
\delta W_Z^{2,\DR} &= \Pi_{WW}^{T,\rm MSSM} - \Pi_{WW}^{T,\rm SM} 
\end{align}
Here, $\Pi_{WW}^{T, \rm MSSM}$ are the full one-loop self-energies within the MSSM. Therefore, 
one needs to  subtract $\Pi_{VV}^{T,\rm SM}$ to include only the new physics contributions. Thus, for consistency, one needs 
to evaluate here $\Pi_{ZZ}^{T,\rm SM}$ in the \DR scheme.   \\

The \DR values of the Weinberg angle and electroweak VEV are now given by
\begin{align}
\sin^2\Theta^{\DR}_W = & \frac{1}{2} - \sqrt{\frac{1}{4}-\frac{\pi \alpha_{ew}^{\DR}(M_{SUSY})} {\sqrt{2} \MZ^2 G_F (1-\delta_r^{\MS} -\delta_r)} } \\
v^{\DR} = & \left(\MZ^{2,\MS}(M_{\rm SUSY})+\delta \MZ^{2,\DR}\right) \frac{(1-\sin\Theta^{\DR}_W) \sin\Theta^{\DR}_W}{\pi  \alpha^{\DR}_{ew}(M_{\rm SUSY}) } 
\end{align}
where the SUSY corrections are calculated as
\begin{align}
\delta_r = &  \frac{1 + \delta \MZ^{2,\DR}/\MZ^2}{1+\delta W_Z^{2,\DR}/m_W^2} \frac{\delta W_Z^{2,\DR}}{m_W^2} - \frac{\delta \MZ^{2,\DR}}{\MZ^2} +  \delta_{VB}^{\DR} \\
\end{align}
$\sin\Theta^{\DR}_W$ and $v^{\DR}$ together with calculated $\alpha^{\DR}_{ew}(M_{\rm SUSY})$ and the input value for $\tan\beta$ determine $g_1^{\DR}(M_{\rm SUSY})$,
$g_2^{\DR}(M_{\rm SUSY})$, $v_d^{\DR}(M_{\rm SUSY})$, $v_u^{\DR}(M_{\rm SUSY})$
\item {\bf Yukawa couplings}\\
As first step, the running $\MS$ Yukawa couplings are translated in $\DR$ values via \cite{Harlander:2006rj}
\begin{align}
m_{e,\mu,\tau}^{\DR}(M_{\rm SUSY}) = & m_{e,\mu,\tau}^{\MS}(M_{\rm SUSY}) \times \left(1-\frac{\alpha_{\rm EW}^{\DR}}{4 \pi} \right) \\
m_{d,s,b}^{\DR}(M_{\rm SUSY}) = & m_{d,s,b}^{\MS}(M_{\rm SUSY}) \times  \Big(1-\frac{\alpha_S^{\DR}}{3 \pi} - \frac{43 (\alpha_S^{\DR})^2}{144 \pi^2}-\frac{\alpha_{\rm EW}^{\DR}}{4 \pi}\frac{1}{9} \Big) \\
m_{u,c,t}^{\DR}(M_{\rm SUSY}) = & m_{u,c,t}^{\MS}(M_{\rm SUSY}) \times  \Big(1-\frac{\alpha_S^{\DR}}{3 \pi} - \frac{43 (\alpha_S^{\DR})^2}{144 \pi^2}-\frac{\alpha_{\rm EW}^{\DR}}{4 \pi}\frac{4}{9} \Big) 
\end{align}
The running Yukawa couplings are obtained from
\begin{equation}
\label{eq:oneloopMF}
m_f^{(1L)}(p^2_i) =  m_f^{(T)} - \tilde{\Sigma}_S(p^2_i) - \tilde{\Sigma}_R(p^2_i) m_f^{(T)} - m_f^{(T)} \tilde{\Sigma}_L(p^2_i) 
\end{equation}
Here, $\tilde{\Sigma}$ are the self-energies without SM contributions. 
The eigenvalues of $m_f^{(1L)}(p^2_i)$ must coincide with $m_{f}^{\DR}(M_{\rm SUSY})$.
\end{enumerate}

\subsubsection{Calculating the \DR parameters at $M_{SUSY}$ in \SPheno}
As in the case of \SARAH,  the \MS parameters are first shifted to \DR parameters and the SUSY thresholds are added
at $Q=M_{\rm SUSY}$. The main difference is, that the conservation of $SU_L(2)\times U_Y(1)$ is assumed at this scale.
The corresponding formulae read as
\begin{enumerate}
\item {\bf Gauge couplings:} these get shifted by
\begin{equation}
\left(g^{\DR}_i\right)^2 = \frac{(g^{\MS}_i)^2}{1 - \frac{(g^{\MS}_i)^2}{8 \pi^2} \Delta g_i^2}
\end{equation}
where
\begin{eqnarray}
\Delta g_1^2 &=& - \sum_{i=1}^3 \left[ \frac{1}{12} \log\left(\frac{m^2_{L_i}}{Q^2}\right)
 +  \frac{1}{12}  \log\left(\frac{m^2_{E_i}}{Q^2}\right)  
 +  \frac{1}{36}  \log\left(\frac{m^2_{Q_i}}{Q^2}\right)  
 +  \frac{1}{18}  \log\left(\frac{m^2_{D_i}}{Q^2}\right) 
 +  \frac{2}{9}  \log\left(\frac{m^2_{U_i}}{Q^2}\right)  \right] \nonumber \\
 &&  - \frac{1}{12} \log\left(\frac{m^2_H}{Q^2}\right)
        - \frac{1}{3}  \log\left(\frac{|\mu|^2}{Q^2}\right) \\
\Delta g_2^2 &=& \frac{1}{3} - \sum_{i=1}^3 \left[ \frac{1}{12} \log\left(\frac{m^2_{L_i}}{Q^2}\right)
 +  \frac{1}{4}  \log\left(\frac{m^2_{Q_i}}{Q^2}\right)  \right] 
 - \frac{1}{12} \log\left(\frac{m^2_H}{Q^2}\right)
 - \frac{1}{3}  \log\left(\frac{|\mu|^2}{Q^2}\right) \nonumber \\
 && - \frac{2}{3}  \log\left(\frac{|M_2|^2}{Q^2}\right) \\
\Delta g_3^2 &=& \frac{1}{2} - \frac{1}{12}\sum_{i=1}^3 \left[ 2\log\left(\frac{m^2_{Q_i}}{Q^2}\right)
 +  \log\left(\frac{m^2_{D_i}}{Q^2}\right) +  \log\left(\frac{m^2_{U_i}}{Q^2}\right)  \right] 
 -  \log\left(\frac{|M_3|^2}{Q^2}\right)
\end{eqnarray}
 and 
 $m_{L_i}$, $m_{E_i}$, $m_{Q_i}$, $m_{D_i}$ and $m_{U_i}$ are the masses of the $\tilde L$, $\tilde E$,
$\tilde Q$, $\tilde D$ and $\tilde U$, respectively, calculated from the corresponding soft SUSY breaking mass
squares. $m_H$ is the mass of the heavy Higgs boson which is calculated according to
\begin{equation}
m^2_H = \frac{1}{2} \left( M^2_{H_u} + M^2_{H_d} + |\mu|^2 + \sqrt{( M^2_{H_u} - M^2_{H_d})^2 + 4 |B \mu|^2} \right)
\end{equation}

\item {\bf Yukawa couplings:}  First the shift from $\MS$ to $\DR$ is calculated according to
\begin{eqnarray}
Y_{SM,l}^{\DR}{}' &=& \left( 1 - \frac{3}{128 \pi^ 2} \left( g^2_1 -  g^2_2 \right) \right) Y_{SM,l}^{\MS} \\
Y_{SM,d}^{\DR}{}'  &=& \left( 1 - \frac{13 g^2_1}{1152 \pi^ 2}  + \frac{3 g^2_2}{128 \pi^2} 
                  - \frac{g^2_3}{12 \pi^ 2}  - \frac{43 g^4_3}{9 (16 \pi^2)^2}   \right) Y_{SM,d}^{\MS} \\
Y_{SM,u}^{\DR}{}'  &=& \left( 1 - \frac{7 g^2_1}{1152 \pi^ 2}  + \frac{3 g^2_2}{128 \pi^2} 
                  - \frac{g^2_3}{12 \pi^ 2}  - \frac{43 g^4_3}{9 (16 \pi^2)^2}   \right) Y_{SM,u}^{\MS} 
\end{eqnarray}
where the gauge couplings $g_i$ are the $\DR$ couplings. In a second step, these
couplings get rescaled as follows
\begin{eqnarray}
Y_{SM,l}^{\DR}{} = \frac{1}{\cos\beta} Y_{SM,l}^{\DR}{}' \,\,,\,\,
Y_{SM,d}^{\DR}{} = \frac{1}{\cos\beta} Y_{SM,d}^{\DR}{}'  \,\,,\,\,
Y_{SM,u}^{\DR}{} = \frac{1}{\sin\beta} Y_{SM,u}^{\DR}{}'
\end{eqnarray}
In the next step, the one-loop corrections due to the SUSY particles
and the heavy Higgs-doublet $H$,
where $H$ is to the SM-Higgs orthogonal combination of $H_u$ and $H_d$. 
Here we distinguish between holomorphic and non-holomorphic corrections where the first
denotes loop contributions to the existing tree-level coupling and the second the loop-induced ones 
to the second Higgs-doublet. 
We give here for simplicity the different contributions for the case of real parameters neglecting
flavour mixing. The case with flavour mixing can be easily obtained from appendix A
Ref.~\cite{Buras:2002vd}. 
\begin{itemize}
\item Taking either $f=t$ or $f=b$ we obtain for the gluino contributions
\begin{eqnarray}
Y_f^{hol} &=& \frac{g^2_3}{6 \pi^2} M_3 T_f C_0(M^2_3,m^2_{Q},m^2_{F}) \\
Y_f^{ahol} &=& -\frac{g^2_3}{6 \pi^2} M_3 Y_f \mu C_0(M^2_3,m^2_{Q},m^2_{F})
\end{eqnarray}

\item Taking either $f=t$, $f=b$ or $f=\tau$ we obtain for the single bino contributions
\begin{eqnarray}
Y_f^{hol} &=& c_f \frac{g^2_1}{16 \pi^2} M_1 T_f C_0(M^213,m^2_{L_f},m^2_{F}) \\
Y_f^{ahol} &=& - c_f \frac{g^2_1}{16 \pi^2} M_1 Y_f \mu C_0(M^2_1,m^2_{L_f},m^2_{F})
\end{eqnarray}
where $L_f=Q$ in case of $f=t,b$ and $L_f=L$ in case $f=\tau$ and the different combinations of
hypercharges give 
\begin{equation}
c_t = -\frac{2}{9} \,\,,\,\, c_b = \frac{1}{9} \,\,,\,\, c_\tau = -1
\end{equation}

\item Taking either $f=t$ or $f=b$ we obtain for the single higgsino contributions
\begin{eqnarray}
Y_f^{hol} &=&  \frac{Y_t Y_b}{16 \pi^2} \mu^2 Y_{f'} C_0(\mu^2,m^2_{Q},m^2_{F'}) \\
Y_f^{ahol} &=& -  \frac{Y_t Y_b}{16 \pi^2} \mu T_{f'}  C_0(\mu^2,m^2_{Q},m^2_{F'})
\end{eqnarray}
where $f'=b$($t$) in case of $f=t$($b$).

\item For the mixed  wino/higgsino contributions we find
\begin{eqnarray}
Y_f^{hol} &=& -\frac{3}{4}  \frac{g^2_2}{16 \pi^2}  Y_{f}  C_2(M^2_2,\mu^2,m^2_{L_f}) \\
Y_f^{ahol} &=& \frac{3}{4}  \frac{g^2_2}{16 \pi^2} \mu M_2 Y_{f}  C_0(M^2_2,\mu^2,m^2_{L_f})
\end{eqnarray}
with $L_f=Q$ in case of $f=t,b$ and $L_f=L$ in case $f=\tau$. 

\item For the mixed  bino/higgsino contributions we find
\begin{eqnarray}
Y_f^{ahol} &=&  - \frac{g^2_1}{16 \pi^2}  Y_{f} 
\left( c_{fL} C_2(M^2_2,\mu^2,m^2_{L_f}) +  c_{fR} C_2(M^2_2,\mu^2,m^2_{F}) \right) \\
Y_f^{ahol} &=&   \frac{g^2_1}{16 \pi^2} \mu M_1 Y_{f} 
\left( c_{fL} C_0(M^2_2,\mu^2,m^2_{L_f}) +  c_{fR} C_0(M^2_2,\mu^2,m^2_{F}) \right)
\end{eqnarray}
with $L_f=Q$ in case of $f=t,b$ and $L_f=L$ in case $f=\tau$. For different coefficients we
obtain
\begin{equation}
c_{tL}= c_{bL} = \frac{1}{6} \,\,,\,\,
c_{tR}=  \frac{2}{3} \,\,,\,\,
c_{bR}=  -\frac{1}{3} \,\,,\,\,
c_{\tau L} = -\frac{1}{2} \,\,,\,\,
c_{\tau R} = 1\,.
\end{equation}

\item Contributions due to the second heavy Higgs doublet with mass $m_H$ read as
\begin{equation}
Y_f^{hol} = c_f \frac{Y_f^3}{16\pi^2} \ln\left(\frac{m^2_H}{M^2_{\rm SUSY}}\right)  
\end{equation}
where $c_f=\sin^2\beta$ in case of $f=b,\tau$ and $c_f=\cos^2\beta$ in case of $f=t$.
\end{itemize}
In case of the $u$-type quarks a simple summation of all contributions suffices
\begin{equation}
Y_u = Y_{SM,u}^{\DR} - \Delta Y_u^{hol} - \Delta Y_u^{ahol} \cot \beta 
\end{equation}
In case of the $d$-type quarks and the leptons one has to resum the aholomorphic contributions as they
get large in case of large $\tan\beta$
\begin{equation}
Y_f = \frac{Y_{SM,f}^{\DR}}{1 + \frac{\Delta Y_f^{ahol}}{Y_{SM,f}^{\DR}} \tan\beta}  - \Delta Y_f^{hol}
\end{equation}
where $f=d,$ 
For completeness we note, that the equivalence of the resummation of the two-point 
function (as done in case of \SARAH) with the resummation of the
three-point function (as done in \SPheno) has been shown in \cite{Carena:1999py}.
\end{enumerate}
The loop functions are given by
\begin{eqnarray}
C_0(m^2_1,m^2_2,m^2_3) &=& \frac{1}{m^2_2 - m^2_3}  
    \left[\frac{m^2_2}{m^2_1 -m^2_2} \ln\left(\frac{m^2_2}{m^2_1}\right)
         - \frac{m^2_3}{m^2_1 -m^2_3} \ln\left(\frac{m^2_3}{m^2_1}\right) \right]\\
C_2(m^2_1,m^2_2,m^2_3) &=& \ln\left(\frac{m^2_3}{M^2_{\rm SUSY}}\right) 
 + \frac	{m_2^4}{(m^2_3-m^2_2)(m^2_2-m^2_1)} \ln\left(\frac{m^2_3}{m^2_2}\right)
 -  \frac{m_1^4}{(m^2_3-m^2_1)(m^2_2-m^2_1)} \ln\left(\frac{m^2_3}{m^2_1}\right)\nonumber\\
\end{eqnarray}

\section{Using the new and old approach in \SARAH / \SPheno}

\subsection{\SARAH}
The new matching routines and Higgs mass calculations are available with \SARAH version 4.9.0. By default, the new routines are included in the \SARAH output of the 
\SPheno source code for any model. Moreover, they are also used by default now for supersymmetric models with the following restriction: 
\SARAH only calculates the effective Higgs pole mass within the SM, if the second lightest CP-even scalar 
has a pole mass above 500~GeV. The reason is that one can expect for lighter mass splitting potential important effects from the mixing between the two lightest scalars which would get lost 
in the effective model ansatz. In addition, there are the following flags which can be used by the user in the LesHouches input file to control when the calculations shall be performed:
\begin{lstlisting}
Block SPHENOINPUT #
...
66  1  # Two-scale matching (yes/no)
67  1  # Calculate Higgs mass in effective SM if possible (yes/no/always)
\end{lstlisting}
The options can be used as follows:
\begin{itemize}
 \item[{\tt 66}]
   \begin{itemize}
   \item[{\tt 0}] the old one-scale matching is used 
   \item[{\tt 1}] the new two-scale matching is used
   \end{itemize}
  The default value is {\tt 1}
 \item[{\tt 67}] 
   \begin{itemize}
   \item[{\tt 0}] the Higgs mass is only calculated at the SUSY scale in the full model
   \item[{\tt 1}] the Higgs mass is calculated in the effective SM if only one light scalar is present   
   \item[{\tt 2}] the Higgs mass is always calculated in the effective SM even if light scalars are present   
   \end{itemize}
 The default value is {\tt 1}
\end{itemize}

\subsection{\SPheno}
\label{app:SPheno}
In \SPheno the new matching procedure and Higgs mass calculation is available with version 4.0.0 and higher.
This procedure is by default switched on but one can switch back to the old one scale matching using the
new entry 49 in block SPHENOINPUT
\begin{lstlisting}
Block SPHENOINPUT #
...
48  1  # 0.. 2-loop QCD to Y_t and alpha_s at m_Z, 1 ... use fit formula at 3 loop
49  1  # Two-scale matching 0/1 correspond to yes/no
\end{lstlisting}
where the value $1$ switches to the one-scale matching. Using 3-loop  fit
formul as given in \cite{Buttazzo:2013uya}
instead of the the 2-loop corrections to
$Y_t^{\MS}$ and 1-loop corrections to $\alpha_s$ at $\MZ$ can be achieved by setting the new
flag 48  in block SPHENOINPUT to 1. 
Moreover, the entry 38 controlling the order used in
the RGEs has been modified 
\begin{lstlisting}
Block SPHENOINPUT #
...
38  3  # 1 & 2: use 1- and 2-loop RGEs; 3: 3-loop SM RGE and 2-loop SUSY RGEs
\end{lstlisting}
with the options   
\begin{itemize}
   \item[{\tt 1}] one loop RGES for both, SM and SUSY
   \item[{\tt 2}] two loop RGES for both, SM and SUSY
   \item[{\tt 3}] three loop RGEs for SM but two loop RGES for  SUSY
\end{itemize}